\documentclass[11pt,a4paper]{article}
\pdfoutput=1

\usepackage{jheppub}

\usepackage{graphicx}
\usepackage{dcolumn}
\usepackage{bm}
\usepackage{amsmath}
\usepackage{amsthm}
\usepackage{amsfonts}
\usepackage{euscript,bbm}
\usepackage{ifthen}
\usepackage{psfrag}
\usepackage{slashed}
\usepackage{mathrsfs}
\usepackage{subfig}

\def\ls{\mathrel{\lower4pt\vbox{\lineskip=0pt\baselineskip=0pt
           \hbox{$<$}\hbox{$\sim$}}}}
\def\gs{\mathrel{\lower4pt\vbox{\lineskip=0pt\baselineskip=0pt
           \hbox{$>$}\hbox{$\sim$}}}}
\def\drawbox#1#2{\hrule height#2pt
\hbox{\vrule width#2pt height#1pt \kern#1pt
              \vrule width#2pt}
              \hrule height#2pt}

\def\Asym#1#2{\vcenter{\vbox{\drawbox{#1}{#2}
              \kern-#2pt       % line up boxes
              \drawbox{#1}{#2}}}}

\newcommand{\be}{\begin{equation}}
\newcommand{\ee}{\end{equation}}
\newcommand{\ben}{\begin{enumerate}}
\newcommand{\een}{\end{enumerate}}
\newcommand{\bei}{\begin{itemize}}
\newcommand{\eei}{\end{itemize}}
\newcommand{\bi}{\begin{itemize}}
\newcommand{\ei}{\end{itemize}}
\newcommand{\mc}{\mathcal}
\newcommand{\Vol}{\mathcal{V}}
\def\nn{\nonumber}

\title{Dark Radiation predictions from general Large Volume Scenarios}

\author{Arthur Hebecker,} 
\author{Patrick Mangat,} 
\author{Fabrizio Rompineve} 
\author{and Lukas T. Witkowski} 

\affiliation{Institut f\"ur Theoretische Physik, Universit\"at Heidelberg, Philosophenweg 19,
D-69120 Heidelberg, Germany
}

\emailAdd{a.hebecker@thphys.uni-heidelberg.de}
\emailAdd{p.mangat@thphys.uni-heidelberg.de}
\emailAdd{f.rompineve@thphys.uni-heidelberg.de}
\emailAdd{l.witkowski@thphys.uni-heidelberg.de}

\abstract{
Recent observations constrain the amount of Dark Radiation ($\Delta N_{\rm eff}$) and may even hint towards a non-zero value of $\Delta N_{\rm eff}$. It is by now well-known that this puts stringent constraints on the sequestered Large Volume Scenario (LVS), i.e.~on LVS realisations with the Standard Model at a singularity. We go beyond this setting by considering LVS models where SM fields are realised on 7-branes in the geometric regime. As  we argue, this naturally goes together with high-scale supersymmetry. The abundance of Dark Radiation is determined by the competition between the decay of the lightest modulus to axions, to the SM Higgs and to gauge fields. The latter decay channel avoids the most stringent constraints of the sequestered setting. Nevertheless, a rather robust prediction for a substantial amount of Dark Radiation can be made. This applies both to cases where the SM 4-cycles are stabilised by D-terms and are small `by accident' as well as to fibred models with the small cycles stabilised by loops. Furthermore, we analyse a closely related setting where the SM lives at a singularity but couples to the volume modulus through flavour branes. We conclude that some of the most natural LVS settings with natural values of model parameters lead to Dark Radiation predictions just below the present observational limits. Barring a discovery, rather modest improvements of present Dark Radiation bounds can rule out many of these most simple and generic variants of the LVS. 
}

\begin{document}
\maketitle

\section{Introduction}
Dark Matter is an essential ingredient of the universe with strong experimental support. While Dark Matter is constituted by hidden non-relativistic particles, there are no reasons \emph{a priori} why hidden sectors could not exhibit species which remain relativistic at CMB and BBN temperatures. Such a sector of particles is called Dark Radiation and it represents an area of Beyond-the-Standard-Model physics, which is increasingly well tested by experiments. The amount of Dark Radiation is conventionally defined as a contribution to $N_{eff}$ -- the effective number of neutrino species. The value of $N_{eff}$ at the CMB time can be measured and any excess $\Delta N_{eff}$ over the Standard Model prediction $N_{eff, SM}=3.046$ is evidence for the existence of Dark Radiation. The most recent result for the effective number of neutrino species is $N_{eff} = {3.30}^{+0.54}_{-0.51}$ (95 \% CL; Planck+WMAP polarisation+high L+BAO) \cite{13035076}. If direct measurements of the Hubble parameter are included the value is modified to $N_{eff} = {3.52}^{+0.48}_{-0.45}$ (95 \% CL; Planck+WMAP polarisation+high L+BAO+${\textrm{H}}_0$) \cite{13035076}.\footnote{Assuming that the recent detection of B-mode polarisation of the CMB by the BICEP2 experiment \cite{14033985} is caused by primordial gravitational waves, a recent analysis \cite{14034852} finds $N_{eff} = 4.00 \pm 0.41$ (68 \% CL; Planck+WP+BICEP2.} The situation is thus unclear at the moment: while there is a mild preference for some Dark Radiation, the absence of Dark Radiation is currently not excluded by experiment. 

Limits on Dark Radiation are powerful tests of Beyond-the-Standard-Model physics and this is particularly true for a popular class of models based on string theory: The scheme of moduli stabilisation known as the Large Volume Scenario (LVS) \cite{0502058} always leads to a Dark Radiation candidate in form of a light axion-like particle (henceforth axion) \cite{12083562, 12083563, 13047987, 14014364}. Dark Radiation is a byproduct of reheating which -- in cosmologies based on string compactifications -- is most naturally caused by the decay of moduli. These dominate the energy density in the post-inflationary period and it is the decay of the lightest modulus which reheats the visible sector but is also responsible for Dark Radiation production. One universal feature of the LVS is the existence of a decay channel for the lightest modulus into light axions, with a decay rate $\Gamma \sim \frac{1}{48 \pi} \frac{m_{\tau_{b}}^3}{M_P^2}$ typical for Planck coupled fields. These axions then constitute Dark Radiation which is thus a generic prediction of the LVS.

In particular, the value of $\Delta N_{eff}$ is controlled by the ratio of the decay rates of the lightest modulus $\tau_{b}$ into light axions versus Standard Model (SM) particles:
\be
\Delta N_{eff} = \frac{43}{7} {\left(\frac{10.75}{g_{*}(T_d)} \right)}^{\frac{1}{3}} {\left. \frac{\rho_{DR}}{\rho_{SM}} \right|}_{T=T_d} = \frac{43}{7} {\left(\frac{10.75}{g_{*}(T_d)} \right)}^{\frac{1}{3}}\frac{\Gamma_{\tau_{b} \rightarrow DR}}{\Gamma_{\tau_{b} \rightarrow SM}} \ ,
\ee
where $T_d$ is the decay temperature of the modulus $\tau_{b}$ and $g_{*}(T_d)$ is the effective number of particle species at that temperature. Thus upper bounds on $N_{eff}$ can be directly interpreted as lower bounds on the decay rates of $\tau_{b}$ into into SM fields. The latter crucially depends on the realisation of the visible sector in the LVS model. 

Dark Radiation has been studied in detail in the sequestered LVS (cf. \cite{12083562, 12083563} in particular), where the visible sector arises from a stack of D3-branes at a singularity.\footnote{In \cite{12083563} the authors study scenarios with the SM at a singularity and with a certain amount of de-sequestering occurring due to moduli-mixing at 1-loop. This scenario differs from the non-sequestered case (with the SM brane in the geometric regime) which we analyse in this work. Moreover, in \cite{12072771} LVS-like racetrack models based on poly-instanton effects have been considered. We comment further on those scenarios when comparing them to our models.} The sequestered LVS has the attractive feature that it allows for TeV soft terms while keeping the gravitino and moduli heavy enough to evade the Cosmological Moduli Problem (CMP) \cite{Coughlan:1983ci, 9308292, 9308325}. However, given the absence of superpartner signals at the LHC, a high SUSY breaking scale may also be an option.\footnote{In addition, the recent measurements of B-modes in the CMB \cite{14033985} may strengthen this point. Assuming that the LVS scalar potential sets the suggested inflation scale of $10^{16}~\mathrm{GeV}$, the compactification volume is expected to be too small to allow for soft masses at $\mc{O}(1)$ TeV in any LVS model. In addition, if the height of the inflaton potential sets the supersymmetry breaking scale, the high scale of inflation suggested by the BICEP2 is an indication for high scale supersymmetry \cite{14036081}.}
      
In the sequestered LVS decays of the lightest modulus into (MS)SM fields are dominated by the decay channel into Higgs scalars: $\tau_{b} \rightarrow H_u H_d$. This coupling originates from a Giudice-Masiero term $ K_{H \bar{H}}(\tau_{b}) (z H_u H_d + c.c.)$ in the K\"ahler potential leading to a decay rate $\Gamma \sim \frac{2 z^2}{48 \pi} \frac{m_{\tau_{b}}^3}{M_P^2}$. All other decay channels into (MS)SM matter are suppressed w.r.t.~this. The upper bounds on $N_{eff}$ then give rise to bounds on this model \cite{12083563}. To arrive at a $N_{eff}$ consistent with experiment one requires a value of $z > 1.5$. Alternatively, this value can be reduced to $z=1$ if one allows for $n_{H}>4$ Higgs doublets. In \cite{13054128} it was shown that these findings are robust if one also considers radiative corrections.

The question thus remains whether it is possible to evade the bounds on $N_{eff}$ in more general constructions of the LVS. In particular, it would be important to determine whether there are realisations of the LVS which are consistent with data on Dark Radiation without the need for additional matter beyond the (MS)SM and for natural values of parameters. 

The purpose of this paper is to examine more general constructions of the LVS and study their predictions for DR. While we aim to be general, we will be most interested in constructions, which maximise the decay rate of $\tau_{b}$ into SM fields compared to the decay rate into light axions and can thus evade stricter bounds on $N_{eff}$. As it turns out, settings where reheating proceeds dominantly through SM gauge bosons predict, for the most natural parameter values, a DR abundance just below present observational bounds.

Let us now step back and examine the various possibilities to boost the decay rate of the lightest modulus into SM fields: 
\begin{enumerate}
\item For one, different realisations of the visible sector could lead to a higher decay rate into matter scalars (squarks, sleptons). These fields couple to the lightest K\"ahler modulus through terms in the K\"ahler potential of the form $K \supset \tau_{b} C \bar{C}$, leading to a decay rate $\Gamma \sim \frac{m_{soft}^2 m_{\tau_{b}}}{M_P^2}$. To make this rate comparable to the decay rate into Higgs fields, we would need a mass for matter scalars close to the threshold $m_{soft} = m_{\tau_{b}} / 2$.
In the sequestered case, soft scalar masses arise at least at a scale $m_{soft} \sim M_P / \mathcal{V}^{2}$, but could also be as high as $m_{soft} \sim M_P / \mathcal{V}^{3/2}$ \cite{09063297}.\footnote{There is further evidence for $m_{soft} \sim M_P / \mathcal{V}^{2}$ from a direct calculation in string perturbation theory \cite{11094153}.} To determine the exact scale requires knowledge of yet undetermined corrections to the K\"ahler potential. The lightest modulus has a mass 
$m_{\tau_b} \sim M_P / \Vol^{3/2}$ and thus, in principle, the decay rate into matter scalars could be comparable to the decay into Higgs fields. However, as we do not know whether $m_{soft} \sim m_{\tau_b}$ can be naturally achieved in variants of the LVS, we will assume that the decay into matter scalars is subleading: $\frac{m_{soft}^2 m_{\tau_{b}}}{M_P^2} \ll \frac{m_{\tau_{b}}^3}{M_P^2}$. In the non-sequestered case, $m_{soft}$ is typically much larger than the mass of the lightest modulus \cite{0505076, 0610129, 10110999} and decays into matter scalars are kinematically forbidden. In consequence, we will not study this decay channel in the following.
\item Further, decays into matter fermions are chirality-suppressed, which is a model-independent statement \cite{12083562}. Hence this decay channel is suppressed w.r.t.~the decay into Higgs fields regardless of the realisation of the visible sector.
\item Last, decays of the lightest modulus into gauge bosons are suppressed in the sequestered LVS, as the modulus $\tau_{b}$ can only decay into visible sector gauge fields at loop level. If there was a tree-level interaction $\tau_{b} F_{\mu \nu} F^{\mu \nu}$ between the modulus and visible sector gauge fields, the decay rate would be increased to $\Gamma \sim \frac{1}{96 \pi} \frac{m_{\tau_{b}}^3}{M_P^2}$, which is comparable to the decay rate into axions or Higgs fields. For this coupling to be present, the visible sector gauge kinetic function has to depend on this modulus. This is the most interesting possibility for increasing the branching ratio of decays of $\tau_{b}$ into visible sector fields and, in the following, we will discuss setups which exhibit this property.
\end{enumerate}

This defines the strategy for this paper. To examine LVS setups which can avoid the most stringent constraints of the sequestered setup, we will explore LVS models, where the lightest modulus reheats the SM by dominantly decaying into gauge bosons. There are two main possibilities:
\begin{enumerate}
\item On the one hand, the lightest modulus could couple to the visible sector gauge bosons directly. 
\item On the other hand, the lightest modulus can decay into gauge bosons which do not belong to the SM gauge group, but under which some SM fields are charged. Such gauge bosons can arise from so-called flavour branes \cite{13040022}, which realise approximate global symmetries of the SM spectrum. The gauge theory on the flavour branes must be broken at some sub-stringy scale where associated gauge bosons become massive. Decays of these gauge bosons then reheat the SM.
\end{enumerate} 

In both cases, for a K\"ahler modulus to couple to gauge bosons at tree-level, we require the cycles supporting the gauge theory on 7-branes to be stabilised in the geometric regime. For case (1.) above, this has immediate consequences for the low energy phenomenology: visible sector cycles in the geometric regime are not sequestered from the source of supersymmetry breaking and, consequently, superpartners typically obtain masses $m_{soft} \sim m_{3/2}$. As $m_{3/2} \gtrsim \mc{O}(10)$ TeV to avoid the CMP these setups necessarily require high scale supersymmetry.\footnote{For example, this situation arises in F-theory GUTs (for reviews see e.g. \cite{12120555, 10093497}) with high-scale SUSY \cite{12062655}.}

However, when coupling $\tau_b$ to the visible sector gauge theory directly, we find the following difficulties. If the cycle supporting the SM gauge group is stabilised supersymmetrically by D-terms, the coupling of $\tau_b F_{\mu \nu} F^{\mu \nu}$ automatically implies a coupling $a_b F_{\mu \nu} \tilde{F}^{\mu \nu}$ of the DR candidate $a_b$ to the SM. In particular, the axion $a_b$ couples to QCD and thus takes over the r\^ole of the QCD axion \cite{PhysRevLett.38.1440, PhysRevD.16.1791}. Consequently, unless there is another axion which can play the r\^ole of the QCD axion, the DR candidate $a_b$ can be identified with the QCD axion for which there are stringent constraints from astrophysics and cosmology (see e.g. \cite{0409059, 0610440, 08071726, 10020329}). For one, we find that our QCD axion candidate produces too much Dark Matter by the vacuum realignment mechanism unless the inital misalignment angle is tuned to $\theta_i \sim 10^{-2}$. Further, if the recent BICEP2 results \cite{14033985} are explained by primordial gravitational waves, the situation is far more severe: we find that setups with $a_b$ as the QCD axion are then ruled out by isocurvature bounds. Similar constraints arise if the cycle supporting the visible sector is stabilised perturbatively by string loop corrections. In this case, there will be an additional light axion beyond $a_b$, which will couple to QCD.

The paper is structured as follows. In section \ref{Sec:Seq} we review DR in the sequestered LVS and examine DR predictions if the requirement of TeV SUSY is lifted. In section \ref{Sec:geo} we study DR predictions for LVS models with visible sectors on D7-branes wrapping cycles in the geometric regime. In particular, we distinguish between setups where ratios of large cycle-volumes are stabilised by D-terms or by loops. 
%-- beyond the LVS procedure -- all cycles are stabilised by D-terms (section \ref{Sec:Dterms}) and setups where both D-terms and string loop corrections are used to stabilise K\"ahler moduli (section \ref{Sec:Loops}). 
Last, we examine DR in LVS models where the SM is reheated via gauge bosons on flavour branes in section \ref{Sec:Flavour}.

\section{Review: Dark Radiation in the sequestered Large Volume Scenario}
\label{Sec:Seq}
In this section we review DR predictions in the sequestered LVS from \cite{12083562, 12083563, 13047987, 13054128}. In particular, we comment on how observational results for $\Delta N_{eff}$ constrain this model and interplay with the SUSY breaking scale.

As in \cite{12083562}, the compactification manifold is taken to be a Swiss-Cheese Calabi-Yau with a volume given by 
\be
\mathcal{V} = \alpha \left( \tau_b^{3/2} - \sum_{i} \gamma_{i} \tau_{s,i}^{3/2}  \right) \ .
\ee
The LVS procedure then fixes $\mathcal{V}$ as well as at least one of the small cycles $\tau_{i,s}$ through an interplay of non-perturbative effects as well as $\alpha^{\prime}$ corrections \cite{0502058}, such that $\mathcal{V} \approx \alpha \tau_b^{3/2}$ is exponentially large. This scheme of moduli stabilisation leads to a clear hierarchy of moduli masses. In particular, the real scalar $\tau_b$ parameterising the bulk volume is the lightest modulus and its axionic partner $a_b$ is essentially massless:
\be
\label{Eq:SeqTmass}
m_{\tau_b} \sim \frac{M_P}{\Vol^{3/2}} \ , \qquad m_{a_b} \sim M_P \ e^{- 2 \pi \Vol^{2/3}} \ .
\ee
Predictions for DR can then be made by studying the decay rates of $\tau_b$ into $a_b$ compared to SM matter.

In the sequestered LVS, the visible sector is realised by D3-branes at a singularity. This scenario is attractive as gaugino and soft scalar masses are suppressed w.r.t.~the gravitino mass. In particular, we follow \cite{12083562} and assume that the suppression is the same for both gauginos and soft scalars:\footnote{The exact scale of soft scalar masses will depend on $\alpha^{\prime}$ corrections to the K\"ahler matter metric which have not been determined yet. Here, we continue with the hypothesis that soft scalar masses only arise at the scale $M_P / \Vol^{2}$. This assumption is strengthened at lowest order in string perturbation theory by a direct calculation \cite{11094153}.}
\be
m_{1/2} \sim m_{soft} \sim \frac{M_P}{\Vol^2} \ll m_{3/2} \sim \frac{M_P}{\Vol^{3/2}} \ .
\ee
Such a hierarchy allows for TeV soft terms while keeping the gravitino and further moduli heavy enough to evade the CMP. The decay rate of $\tau_b$ into SM fields depends on the realisation of the visible sector, which thus introduces a model-dependence into DR predictions.

Given the K\"ahler potential and gauge kinetic function of the low-energy effective theory, the decay rates of the lightest modulus into the visible sector as well as into Dark Radiation can then be computed. For the sequestered LVS the relevant terms are
\begin{align}
\label{Eq:SeqK} K = & \ -3 \ln \left( T_b + \bar{T}_b - \frac{1}{3} \left[C^i \bar{C}^i + H_u \bar{H}_u + H_d \bar{H}_d + \{ z H_u H_d + \textrm{h.c.} \} \right] \right) + \ldots  \\
\nonumber = & \ -3 \ln (T_b + \bar{T}_b) + \frac{C^i \bar{C}^i}{T_b + \bar{T}_b} + \frac{H_u \bar{H}_u + H_d \bar{H}_d}{T_b + \bar{T}_b} + \frac{ z H_u H_d + \textrm{h.c.}}{T_b + \bar{T}_b} + \ldots \ , \\
f_{a} = & \ S + h_{a,k} T_{s_a, k} \ ,
\end{align}
where $T_b=\tau_b+ia_b$ is the bulk volume modulus superfield, $C^i$ are chiral matter superfields and $T_{s_a, k}$ are blow-up modes. %The model-dependence enters via the K\"ahler metric for chiral matter and Higgs fields as well as through the moduli-dependence of the gauge kinetic function.

Given the setting above the rates of decay of $\tau_b$ into DR and SM fields can be determined. One finds that the two dominant decay modes of the volume modulus are the decay into its axionic partner $a_b$ and into Higgs fields. 
\begin{align}
& \textrm{Decays into DR:} & \label{Eq:SeqAA} \Gamma_{\tau_b \rightarrow a_b a_b} &= \frac{1}{48 \pi} \frac{m_{\tau_b}^3}{M_P^2} \ , \\
& \textrm{Decays into SM:} & \label{Eq:SeqHH} \Gamma_{\tau_b \rightarrow H_u H_d} &= \frac{2z^2}{48 \pi} \frac{m_{\tau_b}^3}{M_P^2} \ .
\end{align}
In particular, all other decay channels into visible sector fields are subleading w.r.t.~the decay into $H_u H_d$ as long as $z$ is $\mathcal{O}(1)$. This can be understood as follows: 
\begin{itemize}
\item \emph{Gauge bosons and gauginos}: Decays into gauge bosons are controlled by the moduli-dependence of the gauge kinetic function which, in the sequestered setup, is independent of the light modulus $T_b$ at tree level. Correspondingly, decays of $\tau_b$ into gauge bosons can only occur at loop level, leading to a decay rate which is suppressed by a loop factor: $\Gamma \sim {\left(\frac{\alpha_{SM}}{4 \pi} \right)}^2 \frac{m_{\tau_b}^3}{M_P^2}$. 
\item \emph{Matter scalars:} Decays into matter scalars $C^i$ arise from the term $K_{i \bar{j}} C^i \bar{C}^{\bar{j}}$ leading to a rate $\Gamma \sim \frac{m_{soft}^2 m_{\tau_b}}{M_P^2}$. In the sequestered LVS the soft scale $m_{soft} \sim M_P / \Vol^2$ is parametrically lower than ${m_{\tau_b}} \sim M_P / \Vol^{3/2}$,
thus in turn suppressing the decay rate into matter scalars w.r.t.~\eqref{Eq:SeqHH}.
\item \emph{Matter fermions and Higgsinos:} Starting with \eqref{Eq:SeqK} the decay rate can be determined as $\Gamma  \sim \frac{m_{fermion}^2 m_{\tau_b}}{M_P^2}$, which is chirality-suppressed. Even if decays into fermions are induced at loop level, the decay rate is at most $\Gamma \sim {\left(\frac{\alpha_{SM}}{4 \pi} \right)}^2 \frac{m_{\tau_b}^3}{M_P^2}$, which is again subdominant w.r.t.~\eqref{Eq:SeqHH}.
\end{itemize}

Thus, using \eqref{Eq:SeqAA} and \eqref{Eq:SeqHH} $\Delta N_{eff}$ reads
\be
\Delta N_{eff} = \frac{43}{7} {\left(\frac{10.75}{g_{*}(T_d)} \right)}^{\frac{1}{3}} \frac{\Gamma_{\tau_b \rightarrow DR}}{\Gamma_{\tau_b \rightarrow SM}} = \frac{43}{7} {\left(\frac{10.75}{g_{*}(T_d)} \right)}^{\frac{1}{3}}\frac{1}{n_H z^2} \ ,
\ee
where we also allow for a generic number $n_H$ of Higgs doublets. Clearly, predictions for $\Delta N_{eff}$ also depend on the exact reheating temperature $T_d$ through $g_{*}(T_d)$. This can be determined as $T_d \sim \sqrt{\Gamma_{\tau_b} M_P} \simeq \mathcal{O}(0.1) M_P / \mathcal{V}^{-9/4}$. As the volume $\mathcal{V}$ sets all other scales of the setup, including in particular the supersymmetry breaking scale, the prediction for DR will be discussed in terms of this scale. 

For TeV scale SUSY one finds a reheating temperature of $T_d \lesssim 1$ GeV, which corresponds to $g_{*} =247/4$. We compare the resulting $\Delta N_{eff}$ to the bound $\Delta N_{eff} < 0.79$ (95 \% CL; Planck+WMAP polarisation+high L+BAO) \cite{13035076}. If we allow for one pair of Higgs doublets only, the sequestered LVS is consistent with experimental observation if $z > 1.5$. Alternatively, allowing additional Higgs doublets while fixing $z=1$, the sequestered LVS is not ruled out by observation as long as $n_H > 4$. This requires the field content of the visible sector to be extended beyond the MSSM. 

The above constraints on the sequestered LVS can be somewhat relaxed if one allows for high scale supersymmetry breaking. For example, for $\Vol \lesssim 10^7$, we obtain $m_{soft} \gtrsim 10$ TeV and $T_d \gtrsim 100$ GeV. In this regime we have $g_{*}=106.75$ -- the maximum number in the SM -- and the DR constraints give the following: for $\Delta N_{eff} < 0.79$ and $n_H=2$, we now require $z > 1.3$. If $z=1$, experimental bounds can be met as long as $n_H \ge 4$. While the constraints are less severe, values of $z < 1.3$ still require the addition of matter beyond the MSSM field content. Further, allowing for two Higgs doublets only, small values for the Giudice-Masiero coupling $z< 1$ are still excluded.

It is thus apparent that measurements of $N_{eff}$ impose severe constraints on string models based on the LVS. While there are no a priori reasons why $z > 1$ should be impossible, it remains an open question whether such values can be obtained and whether such a regime occurs naturally in the string landscape. To give an example for the difficulties, note that the particular value $z=1$ can be derived from a shift symmetry in the Higgs sector. In type IIB/F-theory such a symmetry can arise if the Higgs is contained in brane deformation moduli \cite{12042551, 12062655, 13015167, 13042767}. However, in this case the K\"ahler metric is independent of K\"ahler moduli and the lightest modulus $\tau_b$ cannot decay to Higgs fields at all.

\section{Dark Radiation beyond the sequestered Large Volume Scenario}
\label{Sec:geo}

In this section we go beyond the sequestered LVS and analyse DR predictions for more general setups. In particular, while the sequestered LVS considers branes on collapsed cycles, here we examine models with visible sectors given by D7-branes wrapping 4-cycles in the geometric regime. There are in principle three ways how such cycles can be stabilised: 
\begin{enumerate}
\item non-perturbative effects, 
\item gauge-flux-induced D-terms and 
\item string loop corrections to the K\"ahler potential. 
\end{enumerate}
%As non-perturbative effects on the visible sector cycle typically induce a charged superpotential leading to VEVs for %visible matter fields, we do not discuss this possibility in detail. Even if the generation of a charged superpotential is %avoided, we will point out later that such a setup does not lead to an improvement regarding DR limits over the %sequestered case.

\subsection{Visible sector cycle stabilisation by D-terms}
\label{Sec:Dterms}
Consider a Calabi-Yau orientifold $X$ with several 4-cycles $D_i$ whose volumes are given by $\tau_i$. For this discussion it will be useful to also introduce the moduli $t^i$, which give the volumes of 2-cycles. In terms of these the overall volume can be written as $\Vol = \frac{1}{6} k_{ijk} t^i t^j t^k$, where $k_{ijk}$ are the triple intersection numbers of $X$. We also have the relation $\tau_i = \partial \Vol / \partial t^i= \frac{1}{2} k_{ijk} t^j t^k$.

In the following, one ``small'' 4-cycle as well as the overall volume of $X$ will be stabilised using the LVS procedure. All other 4-cycles will be stabilised in a geometric regime by D-terms. In particular, the visible sector will be realised on D7-branes wrapping one of the 4-cycles stabilised by D-terms.

D-terms are induced due to fluxes on D7-branes wrapping these 4-cycles and give rise to a D-term potential 
\be
V_D = \sum_i \frac{g_i^2}{2} {\left( \sum_j c_{ij} {|\phi_j}|^2 - \xi_{i} \right)}^2 \ ,
\ee
where $\xi_i$ are FI-terms and $\phi_j$ are open string states charged under the anomalous $U(1)$ giving rise to the D-term.  
The sum over $i$ is over all 7-branes and the sum over $j$ is over all charged open string states.

We assume that supersymmetric stabilisation can be achieved, without appealing to VEVs of charged fields, by the simultaneous vanishing of all FI-terms: $\xi_i=0$ for all $i$.

The FI-terms are given by an integral over $X$:
\be
\label{Eq:DtermsFI}
\xi_i = \frac{1}{4 \pi \Vol} \int_X \hat{D}_i \wedge J \wedge \mc{F}_i = \frac{1}{4 \pi \Vol} q_{ij} t^j \ ,
\ee
where $\hat{D}_i$ are Poincar\'e dual 2-forms to the 4-cycles $D_i$, $J=t^i \hat{D}_i$ is the K\"ahler form and $\mc{F}_i= \tilde{f_i^j} \hat{D}_j$ is the gauge flux. The $q_{ij}=\tilde{f_i^k} k_{ijk}$ are then the charges of the K\"ahler moduli $T_i$ under the anomalous $U(1)$ \cite{0609211, 11103333}.

From \eqref{Eq:DtermsFI} it is apparent that FI-terms are linear combinations of 2-cycle volumes $t_i$ and hence the requirement $\xi_i=0$ for all $i$ leads to a linear system of equations for the 2-cycle volumes: 
\begin{align}
\label{Eq:DtermsFIsys}
\xi_1 =0 \quad & \Leftrightarrow \quad 0 = q_{11} t^{1} + q_{12} t^{2} + \ldots + q_{1n} t^{n} \ , \\
\nn \xi_2 =0 \quad & \Leftrightarrow \quad 0 = q_{21} t^{1} + \ldots \ . \\
\nn \vdots &
\end{align}  
As a result D-terms fix volumes of some 2-cycles in terms of the volumes of other 2-cycles. 

To combine D-term stabilisation with the LVS, we proceed as follows. We consider the case where one of the 2-cycles, $t^1$, does not appear in the expressions for FI-terms, while all other 2-cycles $t^j$ with $j \neq 1$ are fixed w.r.t.~one another by the system of equations \eqref{Eq:DtermsFIsys}. As a result, $t^1$ remains unfixed at this stage and all other 2-cycles can be expressed in terms of one other 2-cycle, say $t^2$. Further, we consider geometries where $t^1$ enters the volume in a diagonal way: $k_{1jk} = 0$ for $j,k \neq 1$ only, such that $t^1$ only contributes to the volume as $\Vol \supset \frac{1}{6} k_{111} {(t^1)}^3$. Then it follows that
\begin{align}
\tau_1 &= \frac{1}{2} k_{111} {(t^1)}^2 \ , \\
\tau_2 &=  \frac{1}{2} k_{2jk} (t^j t^k) \propto {(t^2)}^2 \ , \\
\textrm{for } i \geq 3: \quad \tau_i &= \frac{1}{2} k_{ijk} (t^j t^k) \propto {(t^2)}^2 \quad \Rightarrow \quad \tau_i = c_i \tau_2 \ ,
\end{align}
where $c_i$ is a numerical factor. On the level of 4-cycles this leads to the desired result: D-terms stabilisation leaves two flat directions which we can parameterise by $\tau_1$ and $\tau_2$. All other 4-cycles are stabilised w.r.t.~$\tau_2$.

Thus, after D-term stabilisation the volume depends on $\tau_1$ and $\tau_2$ only. Here, we consider geometries which lead to a volume of Swiss-Cheese form:\footnote{The fact that both $\tau_1$ and $\tau_2$ appear in the volume in a diagonal way is a direct consequence of the fact that $k_{1jk} =0$ unless $j=k=1$. In this case $\tau_1$ corresponds to a diagonal del Pezzo divisor. It follows that $\tau_1$ has the correct zero-mode structure to give rise to a non-perturbative superpotential.}
\be
\mc{V} = \alpha ( \tau_2^{3/2} - \gamma \tau_{1}^{3/2}) \ .
\ee
WLOG we define $\tau_2 = \tau_b$ and $\tau_1 = \tau_{np}$. The remaining moduli $\tau_b$ and $\tau_{np}$ are then fixed using the standard LVS procedure. Then $\tau_b$ is the lightest modulus as before and its axionic partner $a_b$ is a nearly massless DR candidate (see \eqref{Eq:SeqTmass}). An explicit example for a construction of this type is given in \cite{11103333}.

It behoves to describe what this implies for the visible sector 4-cycle with volume $\tau_a$. While D-terms fix $\tau_a = c_a \tau_b$, the cycle $\tau_a$ has to be fixed small to produce the correct gauge coupling on the visible sector branes: $\alpha_{SM}^{-1} = \langle \tau_a \rangle \approx 25$ (ignoring flux contributions so far). It follows that the parameter $c_a$ has to be tuned such that $\langle \tau_a \rangle$ is small despite $\langle \tau_b \rangle \gg 1$. This amounts to a potentially severe tuning of fluxes, which cannot be avoided in our construction in this section.

As stabilisation by D-terms is achieved at the supersymmetric locus $V_D=0$, the effective theory after D-term stabilisation should still be supersymmetric - i.e.~it should still be formulated in terms of superfields. Thus the condition $\tau_a = c_a \tau_b$ on the visible sector cycle should in fact be enhanced to the condition $T_a = c_a T_b$ at the level of superfields.\footnote{This can also be understood as follows: D-terms not only fix volumes of 4-cycles, they also affect the axion partners. While D-terms stabilise particular combinations of 4-cycle volumes, the same combinations of axions $a_i$ are eaten by the anomalous $U(1)$s and are removed from the low energy theory. As a result, D-terms fix the complete complex moduli $T_i= \tau_i + i a_i$ in terms of other complex moduli. Thus the condition $\tau_a = c_a \tau_b$ on the visible sector cycle should in fact be enhanced to $T_a = c_a T_b$.}

This has the following consequences for the kinetic term of the visible sector gauge theory. Starting with the superfield Lagrangian for the visible sector gauge theory we have:
\be
\label{Eq:DtermsGauge}
\mc{L} \supset \int \textrm{d}^2 \theta \ T_a W_{\alpha} W^{\alpha} = \int \textrm{d}^2 \theta \ c_a T_b W_{\alpha} W^{\alpha} = c_a \tau_b F_{\mu \nu} F^{\mu \nu} + c_a a_b F_{\mu \nu} \tilde{F}^{\mu \nu} \ .
\ee
As a result, the lightest modulus $\tau_b$ now couples to visible sector gauge bosons, which was the objective of the current construction. This opens another channel for $\tau_b$ to decay into SM particles. 

On the other hand, from \eqref{Eq:DtermsGauge} we find that the DR candidate axion $a_b$ now also necessarily couples to the topological term of the visible sector gauge theory including QCD. In this case there will be further constraints on our model. While $a_b$ remains essentially massless after moduli stabilisation, QCD effects will now generate a potential for $a_b$. This changes the cosmological r\^ole of this axion: while a massless axion can only contribute to DR, a massive axion can exhibit both relativistic and non-relativistic populations which can be identified as DR and Dark Matter (DM) respectively. While axions $a_b$ produced by modulus decay will form DR, a population of axion DM will be generated through the misalignment mechanism. The amount of axion DM then crucially depends on the coupling to QCD. The relevant terms in the effective Lagrangian are (see e.g.~\cite{0602233, 12060819}):
\be
\mc{L} \supset K_{T_b \bar{T}_b} \partial_{\mu} a_b \partial^{\mu} a_b + \frac{c_a \tau_b}{4 \pi} F_{\mu \nu} F^{\mu \nu} + \frac{c_a a_b}{4 \pi} F_{\mu \nu} \tilde{F}^{\mu \nu} \ .
\ee
Canonically normalising all fields, the axion couples to QCD as
\be
\label{Eq:DtermsQCDaxion}
\mc{L} \supset \frac{g^2}{32 \pi^2} \frac{c_a a_b}{f_{a_b}} F_{\mu \nu} \tilde{F}^{\mu \nu} \ ,
\ee
where $g^2= c_a \tau_b / 4 \pi$ and $f_{a_b} = \sqrt{K_{T_b \bar{T}_b}} / 2 \pi$ is the axion decay constant. Observational constraints on axion DM are then most conveniently expressed as a bound on $f_{a_b} / c_a$. In particular, observation requires $f_{a_b} / c_a < 10^{12}$ GeV \cite{0610440}. This can be relaxed if the initial misalignment angle is tuned. Alternatively, axion DM could be diluted due to some late time entropy release. There is also a lower bound on $f_{a_b} / c_a$: to avert excessive cooling of stars an axion coupling to QCD has to also satisfy $f_{a_b} / c_a > 10^9$ GeV.

Beyond the gauge kinetic function, the matter K\"ahler metric will also enter the expressions for decay rates of the bulk volume modulus. For matter living on intersections of D7-branes wrapping cycles $\tau_a$, the modulus-dependence of the K\"ahler metric is expected to be of the form $K_{i \bar{j}} \sim \tau_{a}^{1/2} / \tau_{b}$ \cite{0609180} (also see \cite{12062655}). This K\"ahler metric is also appropriate for Higgs fields, as long as they arise from chiral superfields. 

Before calculating the decay rates of the large cycle modulus into visible sector fields, it is worth checking which decay channels are kinematically allowed. In particular, for the non-sequestered setup considered here, the modulus $T_a$ corresponding to the visible sector cycle acquires an F-term and visible sector soft terms are not suppressed w.r.t.~the gravitino mass \cite{10055076, 0610129}. The relevant scales are
\be
\label{Eq:DtermsScales}
m_{soft} \sim m_{3/2} \sim M_P / \mc{V} \qquad m_{\tau_b} \sim M_P / \mc{V}^{3/2} \ .
\ee
Correspondingly, decays of the bulk modulus into matter scalars (and the heavy Higgs) are kinematically forbidden. Thus, in the given setup the volume modulus can only reheat the SM by decaying into gauge bosons and the light Higgs.

There is another observation which can be made from \eqref{Eq:DtermsScales}. To avoid the Cosmological Moduli Problem, we require all moduli masses including $m_{\tau_b}$ to satisfy $m_{mod} \gtrsim 10$ TeV. From \eqref{Eq:DtermsScales} it then follows that $m_{soft} \gg 10$ TeV, and our setup forces us to consider high scale supersymmetry only.

\subsubsection*{Predictions for Dark Radiation}
The decay rates can be calculated from the low energy effective Lagrangian. The relevant terms in the K\"ahler potential and gauge kinetic function are
\begin{align} 
\label{Eq:DtermsK} K = & \ -3 \ln (T_b + \bar{T}_b) + \\ 
\nn & \ + \frac{{(T_a+\bar{T}_a)}^{1/2}}{T_b + \bar{T}_b} \left(H_u \bar{H}_u + H_d \bar{H}_d \right) + \frac{{(T_a+\bar{T}_a)}^{1/2}}{T_b + \bar{T}_b} \left(z H_u H_d + \textrm{h.c.} \right) + \ldots \ , \\
\label{Eq:Dtermsf} f_{a} = & \ T_a + h S \ .
\end{align}

After D-term stabilisation we integrate out $T_a$ by replacing $T_a = c_a T_b$. Rates for decays of the bulk volume modulus can then be determined as
\begin{align}
& \textrm{Decays into DR:} & \label{Eq:DtermsAA} \Gamma_{\tau_b \rightarrow a_b a_b} &= \frac{1}{48 \pi} \frac{m_{\tau_b}^3}{M_P^2} \ , \\
& \textrm{Decays into SM:} & \label{Eq:DtermsHH} \Gamma_{\tau_b \rightarrow hh} & = \frac{z^2}{96 \pi} \frac{\sin^2(2\beta)}{2} \frac{m_{\tau_b}^3}{M_P^2} \ , \\
& & \label{Eq:SeqGG} \Gamma_{\tau_b \rightarrow AA} &= \frac{N_{g}}{96 \pi} \ \gamma^2 \ \frac{m_{\tau_b}^3}{M_P^2} \ , 
\end{align}
where $N_g$ is the number of gauge bosons, $\beta$ is the angle describing the ratio of Higgs VEVs via $\tan \beta$ and we defined
\be
\label{Eq:gamma}
\gamma \equiv \frac{\tau_a}{\tau_a + h \textrm{ Re} (S)} \ .
\ee 
Various values for $\gamma$ correspond to the following regimes. For the case that gauge fluxes do not contribute to the gauge kinetic function ($h=0$) one finds $\gamma=1$. For $| \gamma | \ll 1$ the gauge kinetic function is dominated by the flux-dependent part $h \textrm{ Re} (S)$. For $\gamma \gg 1$ we require a delicate cancellation between contributions from $\tau_a$ and $h \textrm{ Re} (S)$. 

\begin{figure}[t]
 \subfloat[][]{
 \includegraphics[width=0.46\textwidth]{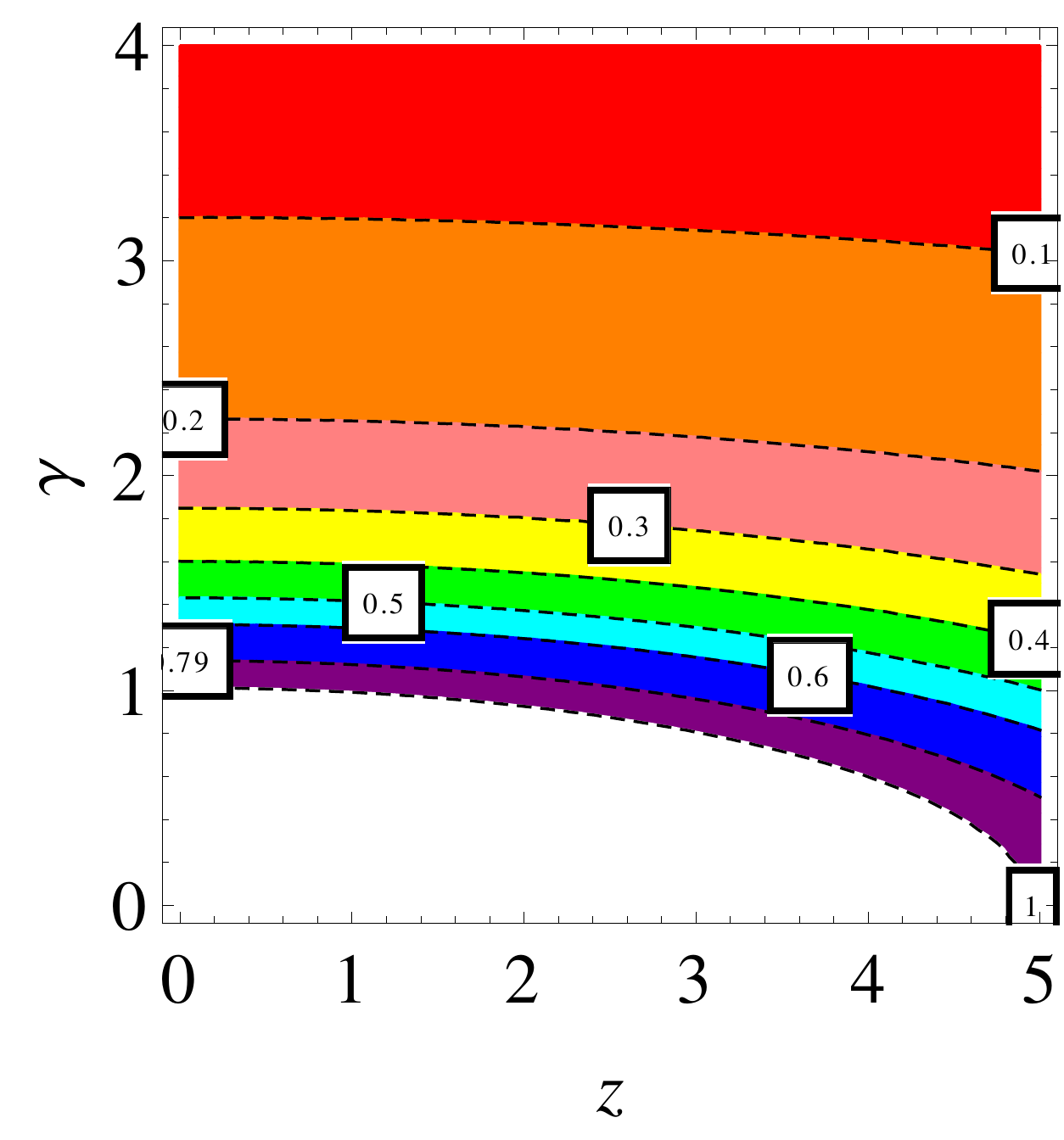}
 }
 \ \hspace{0.5mm} \hspace{5mm} \
 \subfloat[][]{
 \includegraphics[width=0.46\textwidth]{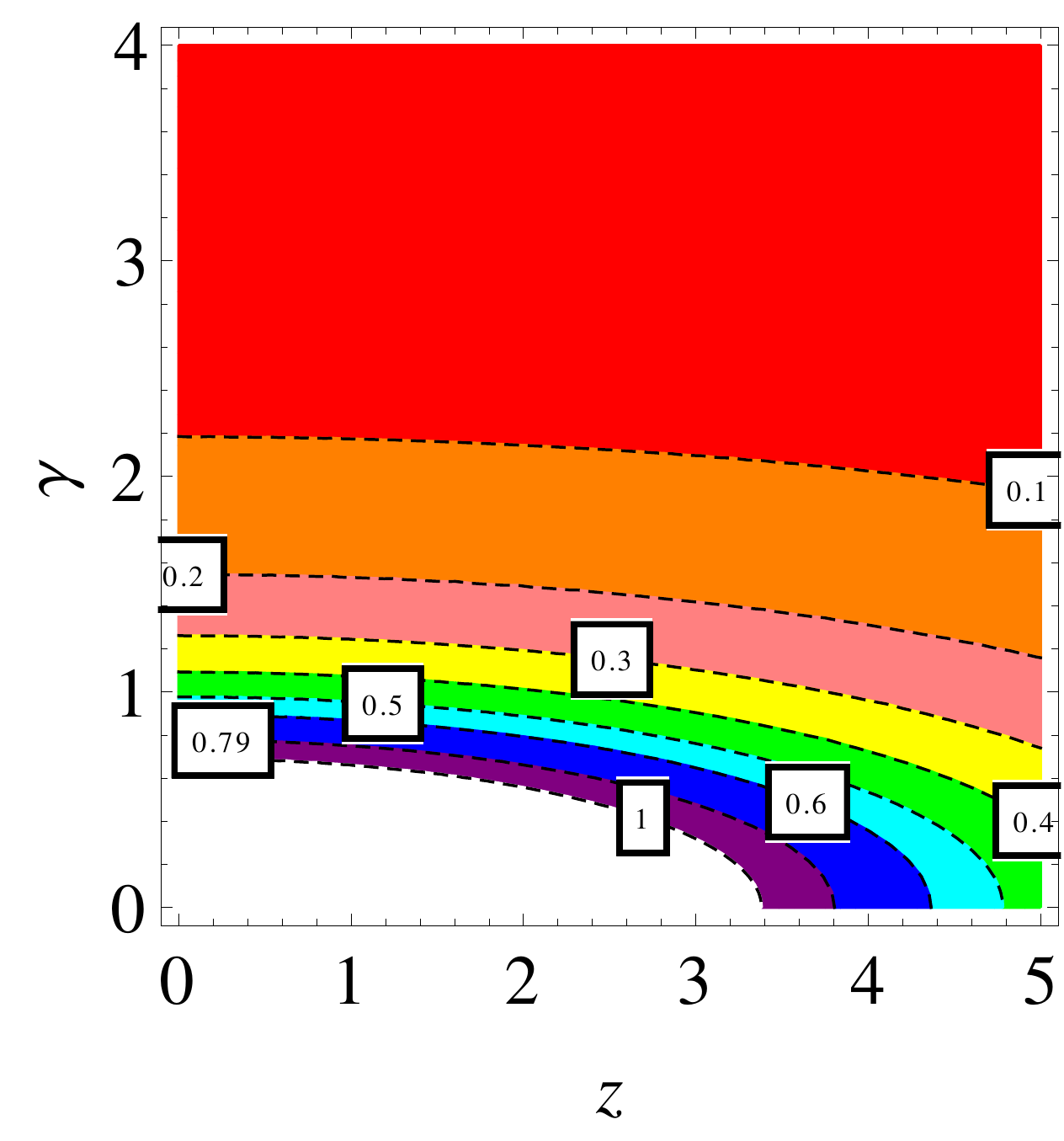}
 }
\caption{Contour plot of $\Delta N_{eff}$ vs.~$z$ and $\gamma$ (defined in \eqref{Eq:gamma}), with (a) $g_{*}=10.75$ and (b) $g_{*}=106.75$. While the predictions are valid for $\sin(2 \beta)=1$, they can be reinterpreted for general values of $\sin(2 \beta)$. To this end we define an effective parameter ${\tilde{z}}^2 = \sin^2(2 \beta) z^2$ and relabel $z \rightarrow \tilde{z}$ on the horizontal axis.}
\label{Fig:Dterms1}
\end{figure}

The decay rate into bulk axions is unchanged relative to the sequestered case \eqref{Eq:SeqAA} while the decay rate into Higgs fields is slightly modified. Using all the above decay rates we obtain the following expression for the amount of DR: 
\begin{equation}
	\Delta N_{\mathrm{eff}} = \frac{43}{7} \left(\frac{10.75}{g_*(T_d)}\right)^{1/3} \frac{\Gamma_{\tau_b \rightarrow DR}}{\Gamma_{\tau_b \rightarrow SM}} =  \frac{43}{7} \left(\frac{10.75}{g_*(T_d)}\right)^{1/3} \frac{1}{\frac{\sin^2(2\beta)}{4}z^2+\frac{N_g}{2} \gamma^2} \ .
\end{equation}
In the following, we set $N_g =12$, which corresponds to the twelve gauge bosons of the SM and we also assume a minimal matter spectrum with only one light Higgs. As we are mainly interested in setups which minimise $\Delta N_{eff}$, we choose $\sin(2 \beta)=1$. This assumptions is further motivated by the fact that high scale supersymmetry suggests $\sin(2 \beta)=1$ \cite{09102235, 12042551, 12056497, 13123235}. The prediction for $\Delta N_{eff}$ shows some mild dependence on $g_*$. We summarised possible values for $g_*$ together with the corresponding reheating temperatures and moduli masses $m_{\tau_b}$ in table \ref{tab:g}. For $g_*=10.75$ and $g_*=106.75$ predictions for $\Delta N_{eff}$ as a function of $z$ and $\gamma$ are shown in figure \ref{Fig:Dterms1} (a) and (b) respectively. 

\begin{table}
\begin{center}
    \begin{tabular}{| r | r | r |}
	\hline
	$m_{\tau_b}$ [GeV] & $T_d$ & $g_*$ \\ 
	\hline
	$5\cdot 10^5$ & $50$ MeV & $10.75$ \\
	$2\cdot 10^6$ & $300$ MeV & $61.75$ \\
	$10^7$ & $3$ GeV & $75.75$ \\
	$\geq 2\cdot 10^8$ & $\geq 200 $ GeV & $106.75$ \\
	\hline
	\end{tabular}
  \caption{Number of degrees of freedom corresponding to different reheating temperatures obtained for different masses of the modulus $\tau_b$. The reheating temperature is determined as $T_d=(1-B_a)^{1/4} {\left({\pi^2 g_*} / {90} \right)}^{-1/4} \sqrt{\Gamma_{\tau_b, total} M_P}$, where $B_a$ is the branching ratio for decays of $\tau_b$ into axions. The values of $T_{d}$ are obtained by considering the very conservative branching ratio $B_a \simeq 0.1$.} 
 \label{tab:g}
\end{center}

\end{table}

Most interestingly, the constraints on $z$ due to bounds on $\Delta N_{eff}$ can be relaxed compared to the sequestered scenario if the modulus decays significantly into gauge bosons. Even if decays into the light Higgs are subdominant, $z \ll 1$, we find that $\Delta N_{eff} < 0.79$ can be satisfied if $\gamma > 0.8$ (for $g_*=106.75$) or $\gamma > 1.2$ (for $g_*=10.75$). Hence, current bounds on DR can be satisfied as long as the gauge coupling is dominantly set by $\tau_a$. 

If bounds on $\Delta N_{eff}$ become stricter in the future, the Higgs sector remains unconstrained if we allow for a mild cancellation between $\tau_a$ and $h \textrm{ Re}(S)$ in the gauge kinetic function. For $\Delta N_{eff} < 0.1$ we require $\gamma > 2.2$ for $g_*=106.75$ ($\gamma > 3.2$ for $g_*=10.75$), corresponding to a fine-tuning between $\tau_a$ and $h \textrm{ Re}(S)$ to $1$ part in $2$ (3).

On the other hand, if decays into gauge bosons are prohibited, we have $\gamma=0$ and restrictions on the moduli-Higgs couplings are even more severe than in the sequestered case.

%Overall we find, that the DR constraints on the LVS are considerably relaxed if the lightest modulus can decay into %gauge bosons. However, for this to be possible, we need the stabilised visible sector cycle to scale with the bulk %volume. In the models considered in this section this was achieved by D-terms which stabilise $\tau_a = c_b \tau_b$. %However, to obtain a visible gauge coupling consistent with observation, the visible sector cycle $\tau_a$ needs to be %hierarchically smaller than $\tau_b$. Here, we need to enforce this condition with an appropriate choice of fluxes, such %that $c_b \sim \frac{25}{\mc{V}}$. Thus for large volumes our setup is severely tuned.   

There are further constraints coming from the fact that $a_b$ becomes massive through its coupling to QCD \eqref{Eq:DtermsQCDaxion} as the axion will contribute to DM through the vacuum realignment mechanism. If the PQ symmetry is broken before inflation, the initial misalignment angle $\theta_i = \frac{a_{b, initial}}{f_{a_b} / c_a} \in [-\pi, \pi)$ is homogeneous in our patch. The axion relic density is then (see e.g.~\cite{0610440}):
\be
\Omega_a h^2 \sim 3 \times 10^3 {\left( \frac{f_a / c_a}{10^{16} \textrm{GeV}} \right)}^{7/6} \theta_i^2 \ .
\ee 
At most the axion density can represents all of cold dark matter, whose density was measured as $\Omega_{DM} h^2 = 0.1199$ \cite{13035076}. Thus, for generic initial misalignment angles there is an overproduction of axion DM if $f_{a_b} / c_a \gtrsim 10^{12}$ GeV. Using \eqref{Eq:DtermsK} we find:
\be
\frac{f_{a_b}}{c_a} = \frac{\sqrt{K_{T_b \bar{T}_b}} M_P}{2 \pi c_a} = \frac{\sqrt{3} M_P}{4 \pi c_a \tau_b} \sim 10^{16} \textrm{ GeV} \quad \textrm{for} \quad c_a \tau_b = \alpha_{vis}^{-1} \sim 25 \ ,
\ee
and thus we only arrive at an acceptable axion DM relic if $\theta_i$ is tuned small. This tuning can be justified anthropically \cite{08071726} and we find that $\theta_i \sim 10^{-2}$ is sufficient to evade DM bounds.

Bounds on isocurvature perturbations can lead to even stricter constraints for the QCD axion. If the measurement of B-modes in the CMB by the BICEP2 experiment \cite{14033985} is explained by primordial tensor modes, the QCD axion candidate with $f_{a_b} / c_a \sim 10^{16}$ GeV will source excessive isocurvature perturbations \cite{0409059}. In consequence, the scenario described in this section would be ruled out.

A possible way out is the existence of another axion $\tilde{a}$ with a decay constant $10^9$ GeV $< f_{\tilde{a}} < 10^{12}$ GeV which couples to QCD:
\be
\label{Eq:DtermsQCDaxion}
\mc{L} \supset \frac{g^2}{32 \pi^2} \left( \frac{\tilde{a}}{f_{\tilde{a}}} + \frac{a_b}{f_{a_b} / c_a} \right) F_{\mu \nu} \tilde{F}^{\mu \nu} \ .
\ee
In this case the QCD axion is mainly given by $\tilde{a}$, which can evade all bounds. For one, it leads to an acceptable DM density without the need to tune $\theta_i$. Further, the decay constant is lower than the Hubble scale of inflation $H_I \sim 10^{14}$ GeV suggested by the BICEP2 results, which implies that the PQ symmetry is intact during inflation. In this case the axion will not source excessive isocurvature perturbations. 
Beyond the QCD axion there will also be a combination of axions, dominantly given by $a_b$, which will remain light and contribute to DR. This axion is then unaffected by isocurvature and DM bounds.

%To summarise, we find that DR constraints on the visible sector in LVS setups can be considerably relaxed, if the %lightest modulus is allowed to decay into SM gauge bosons. However, if this is achieved by stabilising the visible sector %cycle w.r.t.~the bulk volume via D-terms, the setup is severely tuned. First, fluxes need to be adjusted, such that the %visible sector cycle $\tau_a$ needs to be hierarchically smaller than $\tau_b$. In addition, there is the need for further %tuning of the initial misalignment angle of the QCD axion to avert excessive DM production.  

\subsection{Visible sector cycle stabilisation by string loop corrections}
\label{Sec:Loops}

In the previous case we employed D-terms to stabilise the SM branes in the geometric regime. However, to arrive at the correct gauge coupling, this came at the expense of a potentially severe fine-tuning of fluxes. In the following, we will examine scenarios where some cycles are stabilised by string loop effects. 

Explicit examples of LVS models involving string loop stabilisation have been studied for fibred Calabi-Yau manifolds, and it is these spaces which we will analyse in this section. Thus, to be specific, we consider Calabi-Yau three-folds with volume of the form
\begin{equation}
\label{eq:fiberedvol}
\mathcal{V}=\alpha\Big(\sqrt{\tau_{1}}\tau_{2}-\gamma_{np}\tau_{np}^{3/2}\Big).
\end{equation}
If some cycles have been stabilised by D-terms, we assume that this is the volume after the moduli stabilised by D-terms have been integrated out \cite{08080691, 11103333}.  

The overall volume $\mathcal{V}$ and $\tau_{np}$ are stabilised using the standard LVS procedure, such that $\mathcal{V} \cong \sqrt{\tau_1} \tau_2$ is large. There remains one flat direction corresponding of simultaneous changes in $\tau_1$ and $\tau_2$ such that $\mathcal{V}$ is unchanged. It is this mode (denoted by $\chi$ in what follows) which is fixed by string loop corrections as in \cite{08051029}.

To study reheating in this setup we need to specify the visible sector. Here, we model the visible sector by D7-branes on the fiber $\tau_1$. To arrive at an acceptable value for the physical gauge coupling, the volume of the fiber has to be small compared to $\tau_2$ given that $\mathcal{V} \cong \sqrt{\tau_1} \tau_2$ is large. Correspondingly, we need to study the above setup in the ``anisotropic limit'' $\tau_2 \gg \tau_1 \gg \tau_{np}$ also discussed in \cite{08080691, 11103333}. Alternatively, the visible sector could be realised by D7-branes on a further cycle $\tau_a$, whose volume is coupled to the size of $\tau_1$ by D-terms as in \cite{11103333}.\footnote{For a visible sector on two intersecting blow-up modes stabilised by both D-terms and string loops see \cite{12024580}.} In this case the constraints on the volume of $\tau_1$ can be relaxed, as long as the D-term conditions lead to the correct size of $\tau_a$. This corresponds to a tuning of fluxes possibly much weaker than in the previous section. In any case, as the visible sector gauge coupling depends on $\tau_1$ in both cases, our results of this section will be the same for both situations.

Reheating proceeds via decays of the lightest modulus. For the case of the fibred Calabi-Yau considered here there are two moduli which are lighter than all the other closed string moduli. These are the bulk volume and the mode $\chi$ orthogonal to $\mathcal{V}$ with masses
\begin{equation}
m_{\Vol}^2 \sim \frac{M_P^2}{g_s^{3/2} \mathcal{V}^3 \ln \mathcal{V}} \quad \textrm{and} \quad  m_{\chi}^2 \sim \frac{M_P^2}{\mathcal{V}^3 \sqrt{\langle \tau_1 \rangle}}
\end{equation}
respectively \cite{10055076}. If both moduli have comparable masses, we would need to examine the decay of both moduli simultaneously, which complicates the analysis considerably. To simplify the situation, we will thus only consider the case where $\chi$ is lighter than the volume mode, such that the latter can be integrated out. For this to be the case, $\tau_1$ cannot be fixed too small. There is a second reason why $\tau_1$ should not be chosen too small. While $\tau_1$ is a fiber modulus and thus should not give rise to a non-perturbative superpotential of the form $e^{-a T_1}$ due to its zero mode structure, this is not necessarily true in the presence of fluxes. These could lift some of the zero modes such that a non-perturbative superpotential is generated. Thus $\tau_1$ should be large enough that contributions of the form $e^{-a \tau_1}$ in the scalar potential can be safely ignored. 

There is one more point worth mentioning before we study predictions for $\Delta N_{eff}$ in this setup. For every cycle $\tau_i$ that is stabilised by string loop corrections its associated axion $a_i$ will remain light, as perturbative effects cannot generate a potential for axions due to the shift symmetry. Thus, in setups where some cycles are stabilised using string-loop effects, there will be additional light axions which can only contribute positively to $\Delta N_{eff}$. 

Specifically, for the setup considered here, there will be two light axions $a_1= \textrm{Im}(T_1)$ and $a_2= \textrm{Im}(T_2)$. As we saw in the previous section, if one of the axions couples to visible sector gauge fields and, in particular, to QCD, there will be further constraints on this setup. As we realise the visible sector on D7-branes wrapping $\tau_1$, the axion $a_1$ will couple to visible sector gauge fields at tree-level since
\be
\label{Eq:LoopsGauge}
\mc{L} \supset \int \textrm{d}^2 \theta \ T_1 W_{\alpha} W^{\alpha} = \tau_1 F_{\mu \nu} F^{\mu \nu} + a_1 F_{\mu \nu} \tilde{F}^{\mu \nu} \ .
\ee
After canonically normalising all fields we have: 
\be
\mc{L} \supset \frac{g_3^2}{32 \pi^2} \frac{a_1}{f_{a_b}} F_{\mu \nu} \tilde{F}^{\mu \nu} \ ,
\ee 
where $f_{a_1} = {\sqrt{K_{T_1 \bar{T}_1}}} / {2 \pi}$. On the other hand, the axion-like particle $a_2$ will remain light and only contribute to DR. 

\subsubsection*{Predictions for Dark Radiation}
We begin by analysing the decay rates of the lightest modulus $\chi$ into axions, which can be derived from the K\"ahler potential $K= -2 \ln \mathcal{V}$. As there are two light axions in the given case, we present this in some more detail. In particular, we find the following kinetic terms for the moduli $\tau_1$, $\tau_2$ and their corresponding axions $a_1$ and $a_2$:
\begin{align}
\nn \mathcal{L} \supset & \hphantom{+} \frac{1}{4 \tau_1^2} \partial_{\mu} \tau_1 \partial^{\mu} \tau_1 + \frac{1}{2 \tau_2^2} \partial_{\mu} \tau_2 \partial^{\mu} \tau_2 + \frac{\gamma_{np}}{2} \frac{\tau_{np}^{3/2}}{\tau_1^{3/2} \tau_2^2} \partial_{\mu} \tau_1 \partial^{\mu} \tau_2 \ , \\
& + \frac{1}{4 \tau_1^2} \partial_{\mu} a_1 \partial^{\mu} a_1 + \frac{1}{2 \tau_2^2} \partial_{\mu} a_2 \partial^{\mu} a_2 + \frac{\gamma_{np}}{2} \frac{\tau_{np}^{3/2}}{\tau_1^{3/2} \tau_2^2} \partial_{\mu} a_1 \partial^{\mu} a_2 \ .
\end{align}
To arrive at a Lagrangian involving the mode $\chi$ orthogonal to the bulk volume, we integrate out the stabilised volume by setting $\tau_2 = \alpha^{-1} \mathcal{V} \tau_1^{-1/2}$. Up to volume-suppressed terms the fields are then canonically normalised by (see also \cite{10054840})
\begin{equation}
\tau_1 = e^{\frac{2}{\sqrt{3}} \chi} \ , \qquad a_1 = \sqrt{2} a_1^{\prime} \ , \qquad a_2 = \frac{\mathcal{V}}{\alpha} a_2^{\prime} 
\end{equation}
leading to
\begin{equation}
\mathcal{L} \supset \frac{1}{2} \partial_{\mu} \chi \partial^{\mu} \chi + \frac{1}{2} e^{-\frac{4}{\sqrt{3}} \chi} \partial_{\mu} a_1^{\prime} \partial^{\mu} a_1^{\prime} + \frac{1}{2} e^{\frac{2}{\sqrt{3}} \chi} \partial_{\mu} a_2^{\prime} \partial^{\mu} a_2^{\prime} \ .
\end{equation}
In the following, we will drop all primes on the canonically normalised axions. From this Lagrangian the decay rate of $\chi$ into the two light axions can be easily derived. 

The decay rate into Higgs fields is obtained from the following part of the K\"ahler potential: 
\begin{equation} 
\label{Eq:LoopsK} K \supset \frac{{(T_1+\bar{T}_1)}^{1/2}}{\mathcal{V}^{2/3}} \left(H_u \bar{H}_u + H_d \bar{H}_d \right) + \frac{{(T_1+\bar{T}_1)}^{1/2}}{\mathcal{V}^{2/3}} \left(z H_u H_d + \textrm{h.c.} \right) + \ldots \ .
\end{equation}
Similarly, decays into visible sector gauge bosons are determined by the gauge kinetic function
\begin{equation}
\label{Eq:Loopsf} f_{vis} = T_1 + h S \ ,
\end{equation}
where we realised the visible sector by D7-branes wrapping the cycle $\tau_1$.\footnote{For a visible sector given by D7-branes wrapping a different cycle $\tau_a$ the gauge kinetic function has to be modified as $f_{vis} = T_a + h S$. However, if the cycle $\tau_a$ is stabilised w.r.t.~$\tau_1$ via D-terms, such that $T_a = c T_1$, there will still be a direct coupling between $\chi$ and visible sector gauge fields.} Decays into any other Standard Models fields are suppressed as before.

Overall, we find the following decay rates for the modulus $\chi$:
\begin{align}
& \textrm{Decays into DR:} & \label{Eq:LoopsA1A1}  \Gamma_{\chi \rightarrow a_{1},a_{1}}&=\frac{1}{24\pi}\frac{m_{\chi}^{3}}{M_{P}^{2}}\\
& & \label{Eq:LoopsA2A2} \Gamma_{\chi \rightarrow a_{2},a_{2}}&=\frac{1}{96\pi}\frac{m_{\chi}^{3}}{M_{P}^{2}}\\
& \textrm{Decays into SM:} & \label{Eq:LoopsHH}  \Gamma_{\chi \rightarrow h_{1},h_{1}}&=\frac{z^{2}\sin^{2}(2\beta)}{96\pi}\frac{m_{\chi}^{3}}{M_{P}^{2}}\\
& & \label{Eq:LoopsGG} \Gamma_{\chi \rightarrow A_{1},A_{1}}&=\frac{N_{g}}{48\pi}\gamma^{2}\frac{m_{\chi}^{3}}{M_{P}^{2}},
\end{align}
where $\gamma$ is defined as before:
\be
\label{Eq:gamma2}
\gamma \equiv \frac{\tau_1}{\tau_1 + h \textrm{ Re} (S)} \ .
\ee
Thus, for $\gamma=1$, the gauge coupling is dominantly set by $\tau_1$ alone. For $|\gamma| < 1$ the gauge kinetic function is dominated by the flux-dependent part $h \textrm{ Re} (S)$. For $\gamma > 1$ we require a cancellation between contributions from $\tau_a$ and $h \textrm{ Re} (S)$. 

\begin{figure}[t]	
 \subfloat[][]{
 \includegraphics[width=0.46\textwidth]{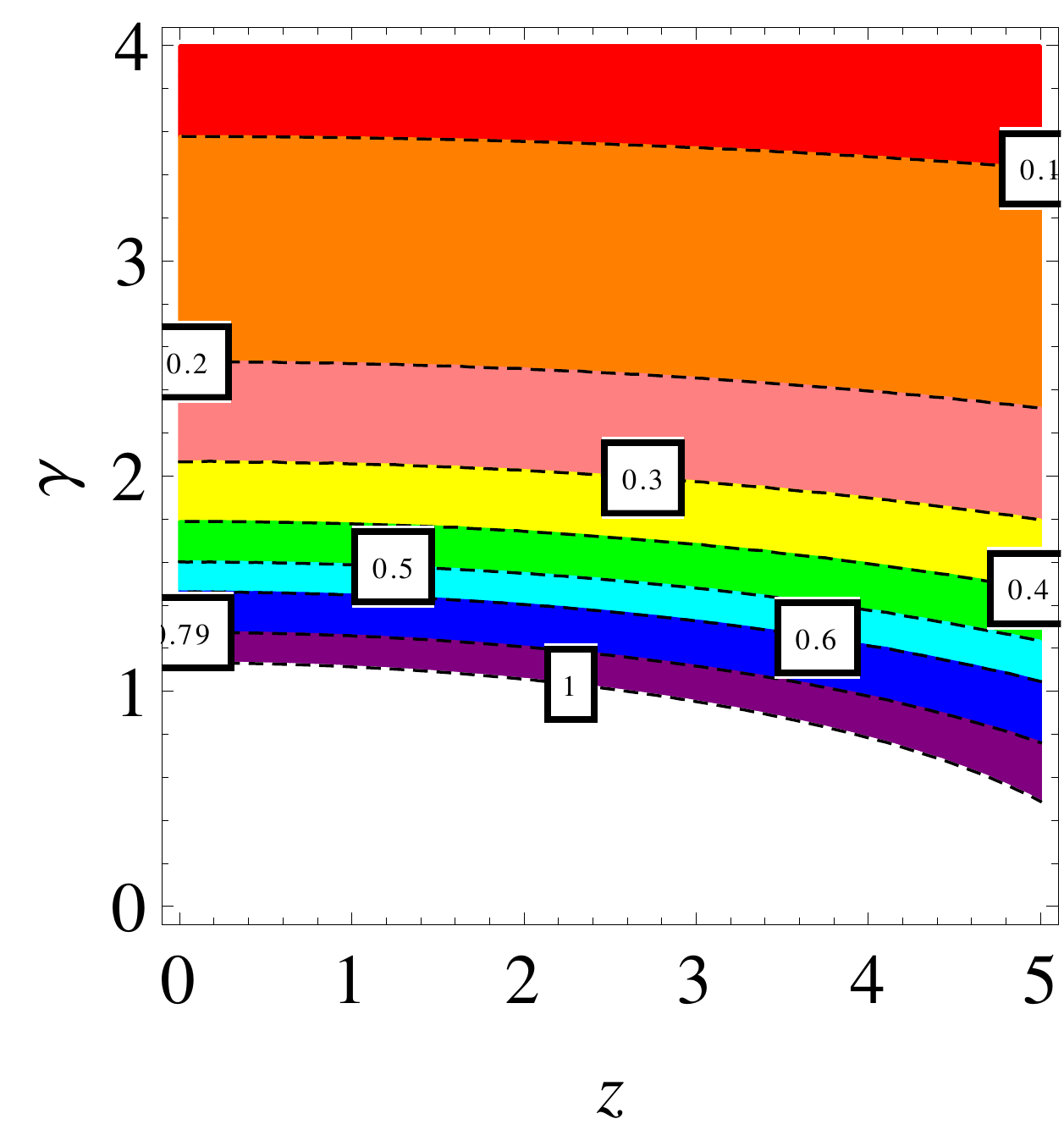}
 }
 \ \hspace{0.5mm} \hspace{5mm} \
 \subfloat[][]{
 \includegraphics[width=0.479\textwidth]{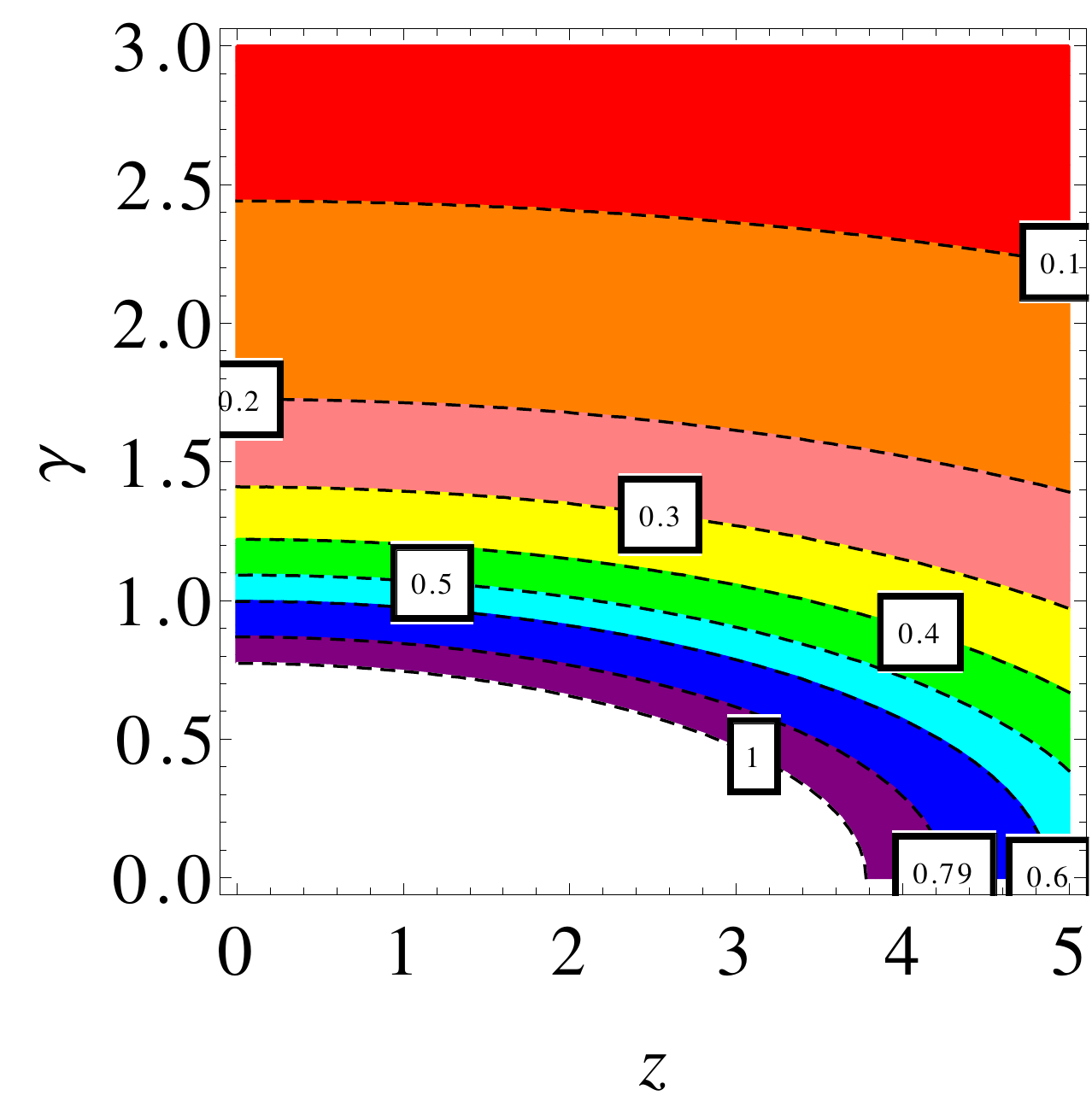}
 }
 \caption{Contour plot of $\Delta N_{eff}$ vs.~$z$ and $\gamma$ (defined in \eqref{Eq:gamma2}) for an LVS model with cycles stabilised by string loop corrections and D-terms, for (a) $g_{*}=10.75$ and (b) $g_{*}=106.75$. As before, the plots have been produced choosing $\sin (2\beta)=1$, but they can be reinterpreted for general values of $\sin(2 \beta)$. To this end we define an effective parameter ${\tilde{z}}^2 = \sin^2(2 \beta) z^2$ and relabel $z \rightarrow \tilde{z}$ on the horizontal axis.}
 \label{Fig:Loops1}
 \end{figure}

Using the above we arrive at the following expression for $\Delta N_{eff}$:
\begin{equation}
	\Delta N_{\mathrm{eff}} = \frac{43}{7} \left(\frac{10.75}{g_*(T_d)}\right)^{1/3} \frac{\Gamma_{\chi \rightarrow DR}}{\Gamma_{\chi \rightarrow SM}} =  \frac{43}{7} \left(\frac{10.75}{g_*(T_d)}\right)^{1/3} \frac{5}{z^{2}+24\gamma^{2}} \ ,
\end{equation}
where we have set $N_g =12$, which corresponds to the twelve gauge bosons of the SM. As before, we also assumed a minimal matter spectrum with only one light Higgs and set $\sin(2 \beta)=1$ to maximise the decay rate into the light Higgs. The effective number of species $g_*$ depends on the reheating temperature and thus on the modulus mass, and it can take values between $g_*=10.75$ and $g_*=106.75$. We present contour plots of $\Delta N_{eff}$ for the two extreme values for $g_{*}$ as a function of $z$ and $\gamma$ in figure \ref{Fig:Loops1}. 

As a result we find that constraints on the Higgs sector from DR can be significantly relaxed if the SM can be reheated via decays of the lightest modulus into gauge bosons. To obtain $\Delta N_{eff} < 0.79$, we require $\gamma > 0.9$ (for $g_*=106.75$) or $\gamma > 1.3$ (for $g_*=10.75$) if all constraints on $z$ are lifted. Consequently, this scenario satisfies current DR bounds when the visible sector gauge coupling is dominantly set by $\tau_1$.

If bounds on $\Delta N_{eff}$ become stricter in the future, the Higgs sector remains unconstrained if we allow for a mild cancellation between $\tau_1$ and $h \textrm{ Re}(S)$ in the gauge kinetic function. For $\Delta N_{eff} < 0.1$ we require $\gamma > 2.5$ for $g_*=106.75$ ($\gamma > 3.6$ for $g_*=10.75$), corresponding to a fine-tuning between $\tau_a$ and $h \textrm{ Re}(S)$ to $1$ part in $3$ (4).

However, as the DR candidate axion $a_1$ couples to QCD, there will also be further constraints. The discussion is analogous to the one in section \ref{Sec:Dterms}, so we do not give details here. In particular, for a generic initial misalignment angle, the vacuum realignment mechanism produces too much axion DM for decay constants $f_{a_1} \gtrsim 10^{12}$ GeV. Here we have:
\be
f_{a_1} = \frac{\sqrt{K_{T_1 \bar{T}_1}} M_P}{2 \pi} = \frac{M_P}{4 \pi \tau_1} \ .
\ee
As the visible sector is realised on $\tau_1$, we require $\tau_1 \sim 25$ for a realistic gauge coupling and the decay constant is $f_{a_1} \sim 10^{16}$ GeV. Thus our setup requires a tuning of the initial misalignment angle 
%to $\theta_i = a_1 / f_{a_1} \sim 10^{-2}$ 
to avert DM overproduction, which can be justified anthropically. 

If the BICEP2 results \cite{14033985} are explained by primordial tensor modes, the situation is far more drastic. In this case the setup described in this section is ruled out, as a QCD axion with $f_{a_1} \sim 10^{16}$ GeV leads to excessive isocurvature perturbations \cite{0409059}. This can be avoided if the model gives rise to an additional axion with a lower decay constant which takes over the r\^ole of the QCD axion as described before in section \ref{Sec:Dterms}.

\subsection{Visible sector cycle stabilisation by non-perturbative effects}
\label{Sec:NP}
We will now describe DR predictions for the case where the visible sector cycle is stabilised by non-perturbative effects. We will find that DR bounds are more restrictive on this scenario than in the sequestered case. Correspondingly, our analysis here will be less detailed than our examinations in the previous sections.

Typically there is a conflict between the presence of both a chiral visible sector and a non-perturbative effect on the same 4-cycle \cite{07113389}. Chiral intersections between the visible sector brane and the non-perturbative effect induce superpotential terms for visible sector fields, which generate VEVs for these fields and break the visible sector gauge theory. However, in \cite{11053193} it was pointed out that D-brane instantons carrying flux can relieve this tension: fluxes can render the instanton superpotential $e^{-T}$ gauge-invariant without the presence of any visible sector fields. Thus it is in principle possible to realise a chiral visible sector on a cycle stabilised by non-perturbative effects.

Here, we will analyse a simple toy model which nevertheless exhibits all the necessary features. To be specific, we consider a compactification with a volume of Swiss-Cheese type:
\be
\Vol = \eta_b \tau_b^{3/2} - \eta_s \tau_s^{3/2} \ .
\ee
Generalisations to setups with more than one cycle of type $\tau_s$ are straightforward. The visible sector will be realised by D7-branes wrapping $\tau_s$. At the same time $\tau_s$ will be wrapped by an E3 instanton or D7-branes exhibiting gaugino condensation, thus giving a non-perturbative contribution $W_{np}$ to the superpotential:
\be
\label{Eq:NPW}
W= W_0 + W_{np}= W_0 +A_s e^{-a_s T_s} \ .
\ee
Here $a_s$ is a model-dependent parameter which depends on the non-perturbative effect wrapping $\tau_s$: For the case of an E3 instanton we have $a_s = 2 \pi$ while for for gaugino condensation on a stack of $N$ D7-branes we have $a_s= \frac{2 \pi}{N}$. The prefactor $A_s$ depends on the dilaton and the complex structure moduli and is a constant at this stage. 

The moduli $\tau_b$ and $\tau_s$ will be stabilised by the standard LVS procedure. As before, the lightest modulus is $\tau_b$ and its axion partner is essentially massless \eqref{Eq:SeqTmass}.

\subsubsection*{Dark Radiation predictions}
Here we will analyse the rates for decays of $\tau_b$ into axions $a_b$ vs.~SM fields. 

For one, realising the visible sector on $\tau_s$ leads to superpartners which are heavier than $\tau_b$: $m_{1/2} \sim m_{soft} \sim M_P / \Vol$ (see e.g.~\cite{0505076}). Hence decays into matter scalars, the heavy Higgs and gauginos are kinematically forbidden. 

As before, decays into the light Higgs will arise through the Giudice-Masiero term with a decay rate of the form 
\be
\Gamma_{\tau_b \rightarrow hh} = \mc{O}(1) \frac{2 z^2}{48 \pi} \frac{\sin^2(2\beta)}{2} \frac{m_{\tau_b}^3}{M_P^2} \ .
\ee
The exact expression for decay rate into the light Higgs will depend on the moduli dependence of the K\"ahler metric for Higgs fields. We do not give more detail as we do not expect any improvement compared to previous sections.

We will be most interested in the decay rate into gauge bosons, which can be calculated given the gauge kinetic function
\be
f_{vis} = T_s + h S \ .
\ee
To study decays of $\tau_b$  we need the effective theory for $\tau_b$ which is obtained by integrating out $\tau_s$. By minimising the F-term potential w.r.t.~$\tau_s$ one obtains (see e.g.\cite{0502058, 12072766}):\footnote{Here, the F-term potential is given by (see e.g.~\cite{12072766})
\be
V= \frac{8 a_s^2 {|A_s|}^2 \sqrt{\tau_s} e^{-2 a_s \tau_s}}{3 \eta_s \Vol} - \frac{4 a_s |A_s| |W_0| \tau_s e^{-a_s \tau_s}}{{\Vol}^2} + \frac{3 \xi \gamma {|W_0|}^2}{4 g_s^{3/2} {\Vol}^3} \ ,
\ee
where $\xi = - \frac{\zeta(3) \chi(X)}{2 {(2 \pi)}^3}$ parameterises the $(\alpha^{\prime})^3$ correction to the K\"ahler potential: $K= -2 \ln (\Vol + \frac{\xi}{2 g_s^{3/2}})$.}
\be
\tau_s = \frac{1}{a_s} \ln \left(\frac{4 a_s A_s}{3 \eta_s} \frac{\Vol}{W_0} \right) + \mc{O}(\ln \tau_s)
\ee
Most importantly, this introduces a dependence of the tree-level visible sector gauge coupling on $\ln \Vol \sim \ln \tau_b$. The rates for the decay channels of interest are then:
\begin{align}
& \textrm{Decays into DR:} & \label{Eq:NPAA} \Gamma_{\tau_b \rightarrow a_b a_b} &= \frac{1}{48 \pi} \frac{m_{\tau_b}^3}{M_P^2} \ , \\
& \textrm{Decays into SM:} & \label{Eq:NPGG} \Gamma_{\tau_b \rightarrow AA} &= \frac{3 N_{g}}{128 \pi} \ \frac{\gamma^2}{(a_s \tau_s)^2} \ \frac{m_{\tau_b}^3}{M_P^2} \ ,
\end{align}
where $\gamma = \frac{\tau_s}{\tau_s + h \textrm{ Re}(S)}$ is defined as in the previous sections, and $N_g$ is the number of generators of the SM gauge group. 

The decay rate into axions is unchanged compared to the sequestered scenario. Further, we find that the decay rate into Higgs fields is not parametrically different from the expressions found in previous sections. In contrast, the decay rate into SM gauge bosons is suppressed by $(a_s \tau_s)^{-2}$ compared to the cases where the visible sector cycle was stabilised by D-terms or string loops. 

It is then easy to see that this construction is more constrained by DR bounds than the setups considered in \ref{Sec:Dterms} and \ref{Sec:Loops}. As the visible sector is realised on $\tau_s$, we require $\tau_s \sim 25$ for an acceptable gauge coupling. In addition $a_s = 2 \pi$ or $\frac{2 \pi}{N}$ depending on the non-perturbative effect sourced by $\tau_s$. Also, unless there is tuning between $\tau_s$ and $h \textrm{ Re}(S)$ we have $\gamma \sim 1$. Then it follows that the decay rate into gauge bosons is typically suppressed w.r.t. the decay rate into axions. This conclusion would only fail if we have $N \gtrsim 100$, corresponding to a gauge group $SU(N \gtrsim100)$ on the stack of D7-branes exhibiting gaugino condensation. Alternatively, one could tune $\gamma$ to adjust the amount of DR. As the amount of DR is not expected to have an effect on the development of life, such a tuning cannot be justified anthropically.

\subsection{Comparison with results from previous work}
Further extensions of LVS models have been suggested in \cite{08062667} and \cite{12072771}, where a non-sequestered LVS is combined with poly-instanton corrections to the superpotential. The authors consider a scenario with a swiss-cheese CY with volume $\mathcal{V} = (\eta_1 \tau_1)^{3/2}-(\eta_2 \tau_2)^{3/2}-(\eta_3 \tau_3)^{3/2}$. Two separate stacks of D7-branes wrapping the 4-cycle $\tau_2$ yield a superpotential with terms $e^{-f_i}$ (race-track model), where the gauge-kinetic function is of the form $f_i=T_2 +C_i e^{-2\pi T_3}+S$ due to gaugino condensations and Euclidean D3-instantons on the non-rigid cycle $\tau_3$. Additionally, the VEV of the flux-superpotential is assumed to be zero.  

This race-track model allows to construct (large volume) minima by integrating out the heaviest modulus $T_2$ near the supersymmetric locus, $\partial_{T_2}W_{np}=0$. Hence, one is left with an effective superpotential $W_{eff}$ which is small due to its exponential suppression by the VEV of $\tau_2$. The stabilisation of $T_1$ and $T_3$ then proceeds as in the usual LVS.    

In our paper we do not consider this scenario any further since no substantial improvement of DR bounds relative to more conventional LVS constructions is expected. The main hope for such an improvement is associated with modulus decays to gauginos which, according to \cite{12072771}, are very light. However, the corresponding rate $\Gamma_{1/2} \sim \frac{M_{1/2}^2m_{\phi}}{M_p^2}$ is much smaller than the decay rate $\Gamma \sim \frac{m_{\phi}^3}{M_p^2}$ of the lightest modulus $\phi$ into axions due to the hierarchy $M_{1/2} \ll m_{\phi}$. This scaling of $\Gamma_{1/2}$ can be understood by expanding the 
generic gaugino Lagrangian
\begin{equation*}
f({\cal V})\lambda\slashed{\partial}\lambda+g({\mathcal{V}})\lambda \lambda\,,
\end{equation*}
where $M_{1/2}=g/f$ is the physical gaugino mass, around the VEV of the volume: ${\mathcal{V}}=\mathcal {V}_0+\delta{\cal V}$. We use the fact that both $f$ and $g$ do not depend on ${\cal V}$ more strongly than through some power, $f\sim {\cal V}^\alpha$ and $g\sim {\cal V}^\beta$. This implies that, at most $f'({\cal V}) \sim f({\cal V})/{\cal V}$, and similarly for $g$. Furthermore, one has to recall that the modulus ${\cal V}$ and the corresponding canonically normalised field $\phi$ are related by $\mathcal{V} \sim \exp(\phi/M_p)$. With the expansion $\phi=\phi_0+\delta\phi$ one then finds the following parametric form of the Lagrangian relevant for the three-particle vertex and hence for the decay:
\begin{equation*}
f({\cal V})\frac{\delta\phi}{M_p}\lambda\slashed{\partial}\lambda+g({\cal V})\frac{\delta\phi}{M_p}\lambda \lambda\,.
\end{equation*}
Now we canonically normalise, $\lambda\to \lambda/\sqrt{f}$, and use equations of motion in the first term, $i\slashed{\partial}\lambda=M_{1/2}\lambda$. This gives a contribution of the order of 
\begin{equation*}
\mathcal{L}_{int} \supset \frac{M_{1/2}}{M_p }\delta \phi \lambda \lambda\,.
\end{equation*}
from both the first and second term above. From here, one can read off the decay rate (up to some numerical factors). Due to its suppression via the mass hierarchy, one can safely neglect this channel.

\section{Large Volume Scenario with flavour branes}
\label{Sec:Flavour}

In the previous sections we examined how the lightest modulus of LVS constructions can be coupled most effectively to the visible sector. Here we will examine the situation where the lightest modulus reheats the visible sector fields via intermediate states. We will argue that gauge bosons arising from the worldvolume theory of so-called flavour branes are ideal candidates for such intermediaries. 

Flavour branes are (stacks of) 7-branes in the geometric regime going through the singularity at which the SM is geometrically engineered. They are known since the very early days of `model building at a singularity' \cite{0005067} and can also be viewed as a tool for generating (approximate) global flavour symmetries of the SM.\footnote{Approximate global symmetries in string theory can also arise from approximate isometries of the compactification space. See \cite{08054037} for more details.} Flavour branes wrap bulk cycles such that for a large bulk volume the gauge theory on their worldvolume is extremely weakly coupled. There will be visible sector states charged under the flavour brane gauge group. This gauge theory has to be spontaneously broken such that, at low scales, a global symmetry of visible sector states emerges. For state-of-the-art string model building employing flavour branes see \cite{13040022}. 

The setup which we are considering in this case is as follows. The Calabi-Yau exhibits a large bulk cycle and a small blow-up cycle giving rise to a non-perturbative effect. These cycles are stabilised by the standard LVS procedure. The visible sector is realised by D3-branes at a singularity as in the sequestered case. However, in addition there are flavour branes, which wrap the bulk cycle but also intersect the singularity. A globally consistent realisation of such a setup in Calabi-Yau orientifolds is described in \cite{13040022}. As we model the visible sector by D3-branes at a singularity, supersymmetry breaking is sequestered and gravity-mediated soft terms are suppressed w.r.t.~the gravitino mass: $m_{soft} \sim M_P / \mathcal{V}^2 \sim m_{3/2} / \mathcal{V}$. However, flavour branes may affect these soft terms. 
%In short, the setup differs from the situation described in \ref{Sec:Seq} and studied in \cite{12083562} simply by the %inclusion of flavour branes.

\subsubsection*{Reheating in LVS models with flavour branes}
We now review the important steps in the cosmological history of the universe, which lead to the reheating of the SM in our setup. 

\begin{enumerate}
\item As in the previous scenarios, the energy density of the universe after inflation is dominated by the lightest modulus, which is the volume modulus $\tau_{b}$ in LVS models\footnote{We do not consider fibred Calabi-Yau manifolds here, where the lightest modulus can be given by a mode orthogonal to the volume.}. In the following, we wish to reheat the visible sector fields via gauge bosons $A_{\mu}$ on flavour branes. If this scenario is to reheat the visible sector more efficiently (and thus lead to a lower $\Delta N_{eff}$) than in the sequestered setup without flavour branes, the decay of $\tau_b$ into pairs of $A_{\mu}$ should be the dominant decay channel of $\tau_b$. In the following we will proceed under this assumption. The decay rate $\Gamma_{\tau_b \rightarrow A_{\mu} A_{\mu}}$ can be determined from the tree-level interaction between $\tau_b$ and $A_{\mu}$ captured by the supersymmetric Lagrangian term $f_{fl} W_{\alpha} W^{\alpha}$, where the gauge kinetic function for flavour branes wrapping a bulk cycle is given by $f_{fl}=T_b$.\footnote{The decay rate $\Gamma_{\tau_b \rightarrow A_{\mu} A_{\mu}}$ depends on $\textrm{Re}(f_{fl})= \tau_b \sim \mathcal{V}^{2/3}$. While there are corrections to the gauge kinetic function due to fluxes such that $f_{fl}= T_b + h S$, these corrections are negligible in here, as $\mathcal{V} \gg 1$.} The resulting decay rate is given by 
\be
\Gamma_{\tau_b \rightarrow A_{\mu} A_{\mu}} = \frac{N_{f}}{96 \pi} \ \frac{m_{\tau_{b}}^3}{M_P^2} \ ,
\ee
where $N_f$ is the number of generators of the flavour brane gauge theory. After $\tau_b$ has decayed the energy density of the universe is then dominated by the gauge bosons $A_{\mu}$ and the axions.
\item The subsequent evolution of the universe then crucially depends on the mass $m_{A}$ of the flavour brane gauge bosons. Hence we will now examine the bounds on $m_{A}$. 
\begin{enumerate}
\item The upper bound on the flavour brane gauge boson mass is given by $m_{A} = m_{\tau_b} /2$, as $A_{\mu}$ are then produced at threshold. To determine the subsequent development of the universe, we determine the decay rate of $A_{\mu}$ into SM particles. In particular, there are SM fermions which are charged under the flavour brane gauge group and the decay rate of $A_{\mu}$ into these fermions is given by 
\be
\label{Eq:Flavourff}
\Gamma_{A_{\mu} \rightarrow f \bar{f}} \sim \alpha_{f} m_{A} \ ,
\ee
where $\alpha_{f} \sim 1/ \tau_b \sim \mathcal{V}^{-2/3}$. One can now easily check that for a wide range of masses $m_{A}$ below threshold the flavour brane gauge bosons decay into SM fields as soon as they are produced by decays of $\tau_b$:
\begin{align}
\Gamma_{A_{\mu} \rightarrow f \bar{f}} & \sim \alpha_{f}^{\prime} m_{A^{\prime}} \sim \mathcal{V}^{-2/3} m_{A^{\prime}} \ , \\ \Gamma_{\tau_b \rightarrow A_{\mu} A_{\mu}} & = \frac{N_{f}}{96 \pi} \ \frac{m_{\tau_b}^3}{M_P^2} \sim \mathcal{V}^{-3} m_{\tau_b} \ .
\end{align}
It follows that $\Gamma_{A_{\mu} \rightarrow f \bar{f}} > \Gamma_{\tau_b \rightarrow A_{\mu} A_{\mu}}$ if $m_{A} > \mathcal{V}^{-7/3} m_{\tau_b}$. The flavour brane gauge bosons then decay into SM fields instantaneously.
\item If $m_{A} \lesssim \mathcal{V}^{-7/3} m_{\tau_b}$, the flavour brane gauge bosons will not decay instantaneously, but form a population of highly relativistic particles carrying a significant fraction of the energy density of the universe. In the end these particles still have to reheat the SM. An interesting question for the following evolution of the universe is whether the flavour brane gauge bosons become non-relativistic before they decay into SM degrees of freedom. If they become non-relativistic, their energy density will scale as matter with time, while any DR produced by the decay of $\tau_b$ earlier will scale as radiation. Consequently, the fraction of the energy density in $A_{\mu}$ over the energy density in DR would grow. As the flavour brane gauge bosons would eventually decay into SM fields, the relic abundance of SM fields would be enhanced with respect to the relic abundance of the axionic Dark Radiation. Correspondingly $\Delta N_{eff}$ could be further suppressed.

However, one can show that the population of $A_{\mu}$ will always remain relativistic until they decay. Initially the population of flavour brane gauge bosons is relativistic with energy density $\rho_{A} = \frac{\pi^2}{30} g_{*, A}(T) T^4$. If the temperature falls to $T \sim m_{A}$ the gauge bosons become non-relativistic. One can now check that $T_{d}$ at which the gauge bosons decay into SM fields is always higher than $m_{A}$. To determine $T_{d}$ we note that $A_{\mu}$ will decay when $\Gamma_{A_{\mu} \rightarrow f \bar{f}} =H$. As the gauge bosons are highly relativistic initially, we need to correct the decay rate into SM fermions by multiplying by a time-dilation factor for relativistic particles. This can be justified \emph{a posteriori}, as we will show that $A_{\mu}$ stay relativistic until they decay. The decay rate \eqref{Eq:Flavourff} is modified as 
\be
\Gamma_{A_{\mu} \rightarrow f \bar{f}} \underset{\textrm{rel.}}{\sim} \alpha_{f} \frac{m_{A}^2}{T} \ ,
\ee
The decay temperature $T_d$ can then be determined using the following equations:
\be
3 H^2 M_P^2 = \rho_{A} = \frac{\pi^2}{30} g_{*, A}(T_{d}) T_{d}^4 \ , \quad H=\Gamma_{A_{\mu} \rightarrow f \bar{f}} \underset{\textrm{rel.}}{\sim} \alpha_{f} \frac{m_{A}^2}{T_{d}} \ ,
\ee
leading to 
\be
T_{d} = {\left( \frac{90}{\pi^2 g_{*,A}} \right)}^{\frac{1}{6}} {\left( \frac{M_P}{m_A} \alpha_{f} \right)}^{\frac{1}{3}} m_{A} \ .
\ee
We recall that for the gauge bosons not to decay instantly when produced their mass had to be small: $m_{A} \lesssim \mathcal{V}^{-7/3} m_{\tau_b} \sim \mathcal{V}^{-23/6} M_P$. It then follows that $T_d > m_A$ and flavour brane gauge bosons always remain relativistic.  

\item While the upper bound on the gauge boson mass $m_A$ is set by the kinematics of the decay of $\tau_b$ we want to examine whether there is a cosmological lower bound on $m_A$. In particular, we will require that when reheating the SM through the decay of flavour brane gauge bosons, the reheating temperature of the SM is $T_{SM} \gtrsim \mathcal{O}(1)$ MeV to allow for standard BBN. To determine the decay temperature of the SM we recall that $A_{\mu}$ will decay into SM fields when $\Gamma_{A_{\mu} \rightarrow f \bar{f}} =H$. We have
\begin{align}
\label{Eq:LoopsH1} 3 H^2 M_P^2 &= \rho_{SM} = \frac{\pi^2}{30} g_{*, SM}(T_{d, SM}) T_{d, SM}^4 \ , \\ 
\label{Eq:LoopsH2}H &=\Gamma_{A_{\mu} \rightarrow f \bar{f}} \underset{\textrm{rel.}}{\sim} \alpha_{f} \frac{m_{A}^2}{T_{d, A}} \ .
\end{align}
To relate $T_{d,A}$ to $T_{d,SM}$ we recall that the comoving entropy density $s= g_{*} a^{3} T^3$ is conserved when $A_{\mu}$ decays:
\be
T_{d,A}= {\left( \frac{g_{*,SM}(T_{d,SM})}{g_{*,A}(T_{d,A})} \right)}^{\frac{1}{3}} T_{d, SM} \ .
\ee
Putting \eqref{Eq:LoopsH1} and \eqref{Eq:LoopsH2} together one finds
\be
m_A = {\left( \frac{\pi^2 g_{*, SM}}{90} \right)}^{\frac{1}{3}} {\left( \frac{g_{*,SM}}{g_{*,A}} \right)}^{\frac{1}{6}} \alpha_{f}^{-1/2} \sqrt{\frac{T_{d,SM}}{M_P}} T_d \ .
\ee
Standard BBN requires $T_d \gtrsim \mathcal{O}(1)$ MeV and thus we find the following lower bound on $m_A$:
\be
m_A \gtrsim \mathcal{V}^{1/3} \sqrt{\frac{1 \textrm{ MeV}}{M_P}} \textrm{ MeV} \ .
\ee
Given that we require $\mathcal{V} \lesssim 10^{14}$ to evade the Cosmological Moduli Problem, it is clear that the cosmological lower bound on $m_A$ is very low. As a result, our setup can successfully reheat the SM -- for a wide range of masses from threshold to the lower limit shown above.

While constraints from reheating allow very light weakly coupled vector bosons there will be further constraints on the parameter space of such particles from collider experiments and precision measurements.
\end{enumerate}
\end{enumerate}

Beyond cosmological constraints, there are also consistency conditions on the string construction. The setup of visible sector and flavour branes has to satisfy local tadpole cancellation conditions. In addition, the local D-brane charges of the flavour branes at the intersection locus with the visible sector have to originate from restrictions of charges of globally well-defined D7-branes. While these consistency conditions do not determine a unique setup of allowed flavour branes, they constrain the number of flavour branes allowed given a particular visible sector \cite{13040022}. 

\subsubsection*{Predictions for Dark Radiation}
Here we determine the decay rates of the lightest modulus into Dark Radiation and Standard Model fields. The lightest modulus is the bulk volume modulus as in the sequestered case or in section \ref{Sec:Dterms}. The rate of decays of $\tau_b$ into its associated axion can be determined from $K=-3 \ln (T_b + \bar{T}_b)$ and gives the familiar result obtained before \eqref{Eq:SeqAA}.

As argued before, decays of $\tau_b$ into gauge bosons on flavour branes lead to a direct reheating of the Standard Model. As flavour branes wrap bulk cycles, there is a tree-level coupling between $\tau_b$ and the gauge bosons on the flavour brane through the kinetic term $f_{fl} W_{\alpha} W^{\alpha}$. For flavour branes on the bulk cycle $f_{fl} = T_b$, where we ignore any flux-induced corrections. Alternatively, we can locate flavour branes on other large cycles $\tau_i$ which intersect the visible sector. The ratio $\tau_i / \tau_b$ is then stabilised by D-terms leading to $f_{fl} = T_i = c T_b$ with $c \sim \mathcal{O}(1)$.

Last, there can also be direct decays of the volume modulus into visible sector matter fields. As described in section \ref{Sec:Seq} the dominating decay channel is given by the interaction of $\tau_b$ with Higgs fields, which arises from the Giudice-Masiero term $(z H_u H_d + \textrm{h.c.})$ in the K\"ahler potential. In contrast to the non-sequestered setups studied before, here both Higgs scalars are light enough to be produced by decays of $\tau_b$ leading to a decay rate \eqref{Eq:SeqHH}.

Overall, we find the following decay rates for the volume modulus:
\begin{align}
& \textrm{Decays into DR:} & \label{Eq:FlavourAA}  \Gamma_{\tau_{b} \rightarrow a_{b},a_{b}}&=\frac{1}{48\pi}\frac{m_{\tau_{b}}^{3}}{M_{P}^{2}} \ , \\
& \textrm{Decays into SM:} & \label{Eq:FlavourGG}  \Gamma_{\tau_{b} \rightarrow A_{\mu}^{flavour} A_{\mu}^{flavour}}&=\frac{N_{f}}{96\pi}\frac{m_{\tau_{b}}^{3}}{M_{P}^{2}} \ , \\
& & \label{Eq:FlavourHH} \Gamma_{\tau_{b} \rightarrow H_u H_d} &= \frac{2z^2}{48 \pi} \frac{m_{\tau_{b}}^3}{M_P^2} \ ,
\end{align}
where $N_f$ is the number of generators of the flavour brane gauge group. 

\begin{figure}[t]	
\centering
\includegraphics[width=0.55\textwidth]{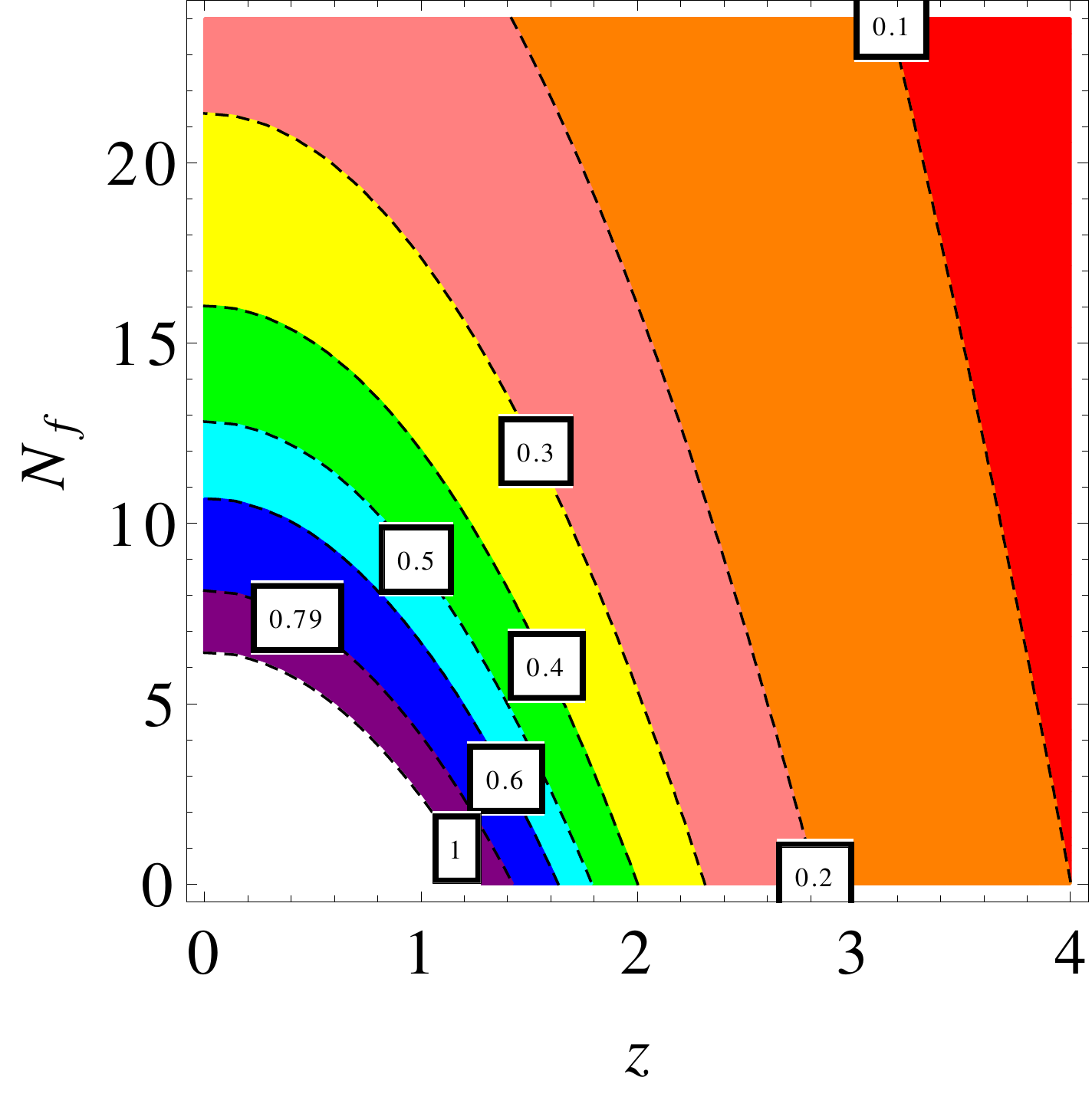}
\caption{Contour plot of $\Delta N_{eff}$ vs.~$z$ and the number of gauge bosons $N_{f}$ of the gauge theory living on the stack of flavour branes wrapping the bulk cycle. The plot was produced for $g_{*}=75.75$, corresponding to a reheating temperature $T_{d}\simeq 1$ GeV.}
\label{Fig:Flavour1}
\end{figure}

Thus we find the following expression for the effective number of neutrino species:
\be
\Delta N_{eff} = \frac{43}{7} {\left( \frac{10.75}{g_{*}(T_d)} \right)}^{\frac{1}{3}} \frac{1}{\frac{N_{f}}{2} + 2 z^2} \ .
\ee
We plot $\Delta N_{eff}$ as a function of $z$ and $N_{f}$, with $g_{*}=75.75$, in figure \ref{Fig:Flavour1}. One finds that the bound $\Delta N_{eff} < 0.79$ can be achieved without any restrictions on the Higgs sector as long as $N_f \ge 9$. For $z=1$ the DR bound $\Delta N_{eff} < 0.79$ requires at least $N_{f} \ge 5$ gauge bosons. Thus for a number of flavour branes as small as $10$ current bounds on DR can be easily met. 

\begin{table}
\begin{center}
    \begin{tabular}{| r | r | r |}
\hline
$\Delta N_{eff}$ & \multicolumn{2}{c|}{$N_{f}$} \\ \cline{2-3}
& $z=0$ & $z=1$ \\ \hline
$< 0.79$ & $> 9$ & $>5$ \\
$< 0.50$ & $> 13$ & $>9$ \\
$< 0.40$ & $> 16$ & $>12$ \\
$< 0.30$ & $> 22$ & $>18$ \\
$< 0.20$ & $> 32$ & $>28$ \\
$< 0.10$ & $> 65$ & $>61$ \\ \hline
    \end{tabular}
\end{center}
\caption{Minimum number of gauge bosons $N_f$ on flavour branes needed to evade upper bound on $\Delta N_{eff}$ for $z=0$ and $z=1$} \label{tab:flavour1}
\end{table}

If bounds on $\Delta N_{eff}$ become more restrictive, we have the following options. If we do not wish to impose constraints on the Higgs sector of the model, we require a larger number of generators on flavour branes. A table of the minimum numbers of gauge bosons needed for fixed $z$ given an upper bound on $\Delta N_{eff}$ is shown in \ref{tab:flavour1}. However, conditions for globally consistent models restrict the maximum number of flavour branes for a given visible sector \cite{13040022} and thus place an upper limit on the allowed number of gauge bosons. The exact constraints on the numbers of flavour branes will depend on the details of the individual model. While we cannot be more specific it follows that lower bounds on $\Delta N_{eff}$ cannot necessarily be evaded by simply introducing more flavour branes. 

%To summarise, we find that LVS setups with not more than $n=10$ flavour branes can evade current DR bounds.

\section{Conclusion}
In this paper we examined predictions for the amount of Dark Radiation (DR) for string models employing the scheme of moduli stabilisation known as the Large Volume Scenario (LVS). By analysing the ratio of decays of the lightest modulus into axions vs.~Standard Model particles in these setups we study contributions to the effective number of neutrino species $\Delta N_{eff}$ produced during reheating. 

We find that DR bounds on LVS models are considerably relaxed if the lightest modulus can reheat the Standard Model by decaying into gauge bosons. We consider setups where the modulus couples directly to visible sector gauge bosons and we also examine models where the modulus decays into gauge bosons on flavour branes which subsequently reheat the Standard Model. 

In the first case we find that models can evade current DR bounds for natural values of parameters. In particular, we find $\Delta N_{eff} < 0.79$ as long as the visible sector gauge coupling ${4 \pi}/{g^2}= \tau_{vis} + h \textrm{ Re}(S)$ is dominantly set by $\tau_{vis}$. However, if bounds on DR become stricter in the future, the parameter space is increasingly constrained. In particular, the amount of DR can be reduced if one allows for a cancellation between $\tau_{vis}$ and $h \textrm{ Re}(S)$. For a mild tuning of 1 part in 2 between $\tau_{vis}$ and $h \textrm{ Re}(S)$ the amount of DR can be reduced to $\Delta N_{eff} \sim 0.1 - 0.2$.

We also observe that when coupling the lightest modulus $\Phi$ to visible sector gauge bosons, the model is necessarily non-sequestered and $m_{soft} \gg m_{\Phi}$. As we require $m_{\Phi} > \mc{O}(10)$ TeV to solve the Cosmological Moduli Problem we have $m_{soft} \gg \mc{O}(10)$ TeV, so that we are led to a regime of high scale supersymmetry.

In addition, by coupling the lightest modulus -- a saxion -- to visible sector gauge bosons, the DR candidate axion is coupled to the topological term of the visible sector gauge theory and takes the r\^ole of the QCD axion. In particular, the axion $a$ couples to QCD as
\be
\mc{L} \supset \frac{g^2}{32 \pi^2} \frac{a}{(M_P / 4 \pi \tau_{vis})} F_{\mu \nu} \tilde{F}^{\mu \nu} \sim \frac{g^2}{32 \pi^2} \frac{a}{\mc{O}(10^{16}) \textrm{ GeV}} F_{\mu \nu} \tilde{F}^{\mu \nu} \ ,
\ee
leading to an overproduction of axion Dark Matter through the misalignment mechanism. This can be averted if the initial misalignment angle is tuned to $\theta_i \sim 10^{-2}$. 

Further, if the recent BICEP2 results \cite{14033985} are explained by primordial gravitational waves setups with $a$ as the QCD axion are ruled out by isocurvature bounds. A possible way out is the existence of an additional axion with a decay constant below the scale of inflation which can take over the r\^ole of the QCD axion.

For models with flavour branes, we find that current DR bounds ($\Delta N_{eff} < 0.79$) can be satisfied if flavour branes give rise to $N_f = 5 - 9$ gauge bosons. This scenario requires that the gauge theory on the flavour branes is broken in such a way, that the gauge bosons stay light enough to be produced by decays of the lightest modulus. The DR axion does not couple to QCD in this case and there is no immediate problem with overproduction of axion DM or isocurvature bounds. However, a realistic model would require an additional axion which would take the r\^ole of the QCD axion.

One possibility to further enhance the decay rate of the lightest modulus to the SM and thus to evade possible stronger DR bounds is the following: Recall that the special r\^ole of the Higgs in the decays to the SM arises because the supersymmetric Higgs sector allows for a Giudice-Masiero type term in the K\"ahler potential. Such a term cannot arise for other SM fields due to chirality. Instead of duplicating the Higgs sector, one might consider singlets, which come naturally in many string constructions and can also play the r\^ole of right-handed neutrinos. Allowing for many such fields with appropriate Giudice-Masiero terms and couplings to Higgs and lepton-doublets has the potential to naturally open up further decay channels to the SM. 

Overall, Dark Radiation is a powerful tool to constrain string models of particle physics based on the LVS. Moreover, as we have seen in great detail, some of the most natural settings with natural values of model parameters lead to Dark Radiation predictions just below the present observational limits.

\textit{This paper was submitted simultaneously to the related work \cite{14036473}.}

\subsection*{Acknowledgments}
We thank Joseph Conlon, M.C.~David Marsh, Michele Cicoli, Joerg Jaeckel, Stephen Angus, Eran Palti and Timo Weigand for interesting conversations and helpful comments. This work was supported by the Transregio TR33 ``The Dark Universe''.

\bibliographystyle{JHEP}
\bibliography{DR2bib}

\end{document}